\documentclass[12pt]{elsarticle}
\usepackage[a4paper, total={7in, 10in}]{geometry}

\usepackage{hyperref}
\usepackage{ tipa }
\usepackage{xcolor}
\usepackage{tipa}
\usepackage{graphicx}
\usepackage{subfig}
\usepackage{amsmath}
\usepackage{nomencl}
\usepackage{comment}
\makenomenclature

\begin{document}

\begin{frontmatter}

\title{How does human blood rheology influence arterial and cardiovascular device hemodynamics? A comprehensive review, discussion, and future directions}

\author{C. Sasmal\footnote{Email: csasmal@iitrpr.ac.in (corresponding author)}}
\address{Department of Chemical Engineering, Indian Institute of Technology Ropar, Rupnagar, India-140001}





\begin{abstract}
 Human blood is a complex biological suspension whose macroscopic flow behaviour arises from the coupled microscopic dynamics of deformable blood cells, plasma proteins, and evolving cellular microstructures. As a result, blood exhibits a range of nonlinear rheological behaviours, including shear-thinning, viscoplasticity, viscoelasticity, and thixotropy, which strongly influence its physiological and clinical hemodynamics. This review provides a comprehensive assessment of the role of blood rheology in arterial and blood-contacting cardiovascular device flows, with particular emphasis on the extent to which different rheological characteristics alter clinically relevant flow quantities and on whether the conventional Newtonian approximation remains adequate for the analysis. The underlying physical mechanisms responsible for the major rheological behaviours of blood are first discussed, followed by a critical review of the constitutive models commonly employed in hemodynamic analysis, including generalised Newtonian, viscoelastic, and thixo-elasto-viscoplastic (TEVP) ones. The effects of blood rheology on arterial flows involving stenosis, bifurcation, and aneurysm under both steady and pulsatile flow conditions are subsequently reviewed and discussed in terms of several physiologically important parameters, such as pressure drop, flow resistance, wall shear stress, flow separation, and recirculation zone. The influence of blood rheology on cardiovascular devices, including prosthetic heart valves and stents, is also reviewed, particularly regarding flow stagnation, residence time, hemolysis, and thrombosis. This review highlights that shear-dependent viscosity is generally the most established and widely used rheological effect in hemodynamic modelling, whereas the importance of yield stress, viscoelasticity, and thixotropy has not yet been systematically investigated to the same extent. Importantly, these rheological behaviours should not necessarily be considered independently, since they can coexist and interact under spatially heterogeneous flow conditions arising because of pulsatile flow in cardiovascular systems. In this context, thixo-elasto-viscoplastic models offer a physically comprehensive framework for describing the coupled evolution of blood microstructure, elastic stress, yielding, and viscous response, although their additional computational complexity remains substantial. Finally, key research gaps are identified, including systematic model-to-model comparisons under identical flow conditions and experimental validation using spatially and temporally resolved flow fields. Such efforts are essential for establishing when the additional complexity of blood-specific rheological modelling provides meaningful improvements over Newtonian and generalised Newtonian descriptions, and for developing more reliable tools for cardiovascular disease assessment and medical device design.    

\end{abstract}


\end{frontmatter}


\section{Introduction}
\begin{figure}
    \centering
    \includegraphics[width=10cm]{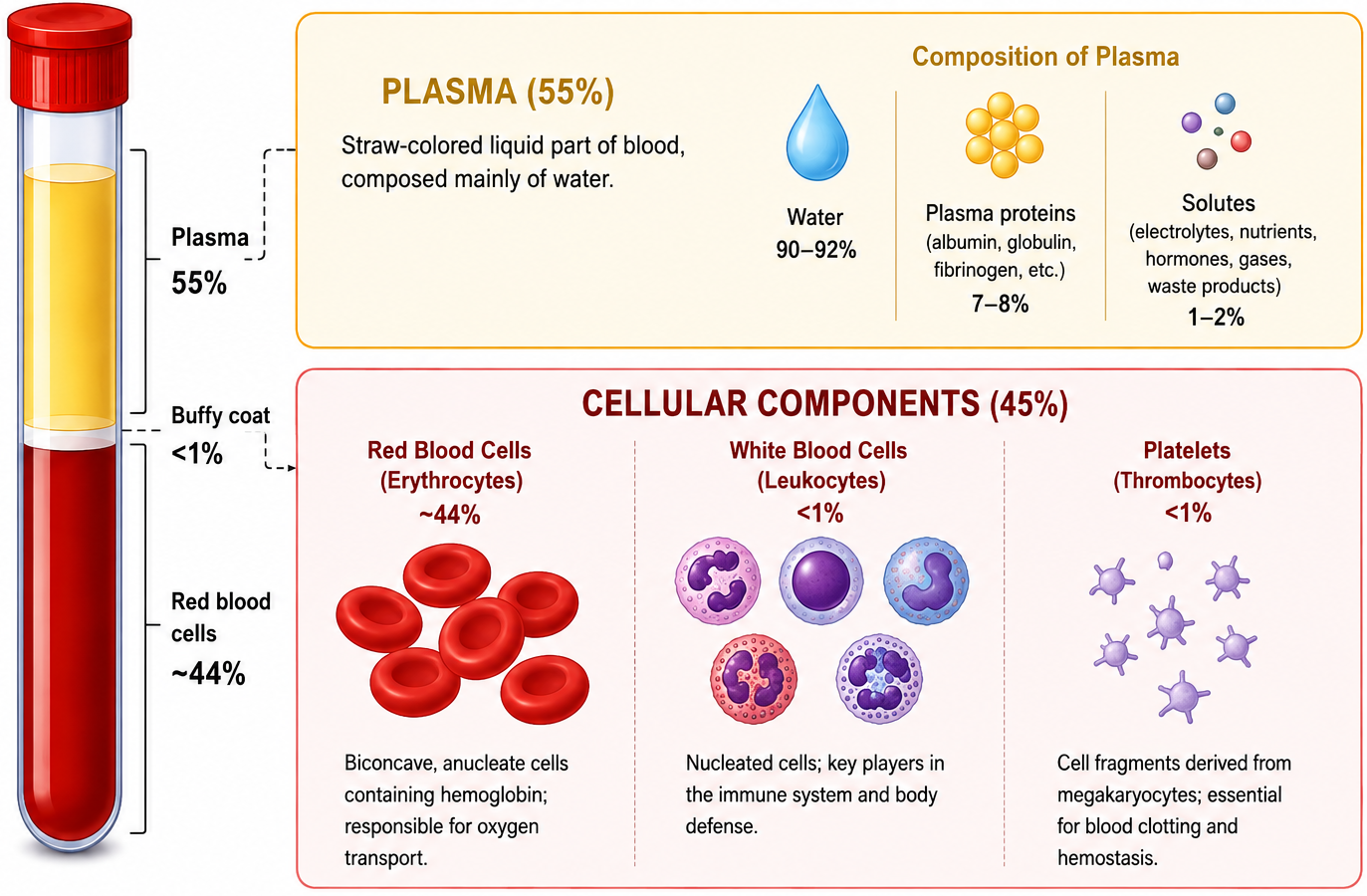}
    \caption{Schematic of different human blood components~\cite{cooper2016blood}.}
    \label{Blood}
\end{figure}
Human blood is a multicomponent biological fluid that accounts for 6-8\% of body weight and is essential for survival and the maintenance of physiological functions. It consists of two phases, namely, a liquid phase and a cellular phase. The liquid phase is known as the plasma phase (about 55\% of the total blood), which contains mostly water (90-92\% of the total plasma), along with proteins such as albumin, globulins, and fibrinogen, several electrolytes, nutrients, hormones, and dissolved gases. On the other hand, the cellular phase (about 45\% of the total blood) consists of red blood cells (erythrocytes), white blood cells (leukocytes), and platelets~\cite{cooper2016blood}. Each of these different components of blood has distinct activities, as schematically shown in Figure~\ref{Blood}. For instance, red blood cells (RBCs) transport oxygen and carbon dioxide, white blood cells (WBCs) provide immune defence against pathogens, platelets facilitate blood clotting and wound healing, and plasma acts as a transport medium so that cells get what they need and waste is removed efficiently~\cite{schaller2008human,sturkie1986body}. Apart from transport, blood also regulates body temperature, maintains pH, and facilitates communication among organs through hormones and other signalling molecules. Because blood interacts with almost every organ system in our body, understanding its flow behaviour is fundamentally vital for effectively diagnosing diseases, performing medical treatments, and developing blood-contacting medical devices. 

Therefore, it can be readily acknowledged that human blood is not a simple fluid like water, but a dense suspension of deformable and interacting constituents. As a result, when a force is applied to blood, it does not exhibit a simple linear relationship between the applied force and deformation rate (i.e., Newton's law of viscosity); instead, it displays a range of complex nonlinear relationships, mostly induced by the dominant cellular component RBCs, which are highly flexible biconcave structures. At low deformation or shear rates, they tend to form aggregates, known as rouleaux, which increase blood viscosity~\cite{baumler1999basic,chien1973ultrastructural}. However, as the deformation rate increases, these aggregates are progressively broken, and RBCs align along the flow direction, reducing resistance to flow and decreasing blood viscosity, resulting in shear-thinning rheological behaviour~\cite{beris2021recent}. The presence of plasma proteins such as fibrinogen and globulins facilitates intercellular adhesion, and a finite force is required to disrupt these microstructural networks before flow begins, resulting in viscoplastic or yield-stress behaviour in blood~\cite{picart1998human,merrill1969yield,morris1989evaluation}. Furthermore, the elastic membranes of RBCs impart elasticity to blood by storing and releasing deformation energy, which is another complex nonlinear rheological behaviour exhibited by human blood~\cite{puig2007viscoelasticity,thurston1972viscoelasticity}. The formation of RBC aggregates and their progressive breakdown under deformation also lead to the generation of more complex, history-dependent viscosity, or thixotropic rheological behaviour, in human blood~\cite{huang1975quantitative,huang1987thixotropic,javadi2022thixotropy}. Even blood plasma alone without the cellular constituents also exhibits nonlinear viscoelastic behaviour~\cite{brust2013rheology}. Although present at lower volume fractions, the additional contributions from WBCs and platelets are also involved in these transient interactions and promote these various complex rheological behaviours of blood. Altogether, these multiscale interactions among deformable blood cells, plasma proteins, and an evolving microstructure make it inevitable that blood exhibits rich, non-Newtonian rheological behaviours rather than behaving like a simple fluid. Some excellent review articles on this topic are available for readers to look into~\cite{beris2021recent,fatahian2018review,stuart1990technological,nader2019blood}. \\      

Due to these complexities in the rheological behaviour of blood, one can expect that its flow behaviour, or hemodynamics, would also be more complex and richer in physics than when assuming blood as a simple Newtonian fluid. Therefore, several clinically important quantities, directly obtained from hemodynamics, such as wall shear stress (WSS), pressure drop, flow separation, and residence time, are expected to be influenced by these nonlinear rheological behaviours. For instance, the shear-thinning of blood causes its viscosity to be varied spatially with local deformation or shear rate. In low-shear regions, such as arterial bifurcations, aneurysms, or recirculation zones in blood-contacting medical devices, blood viscosity increases significantly due to RBC aggregation. In contrast, near the arterial wall, blood viscosity decreases due to high shear. This directly alters the WSS distributions, a key physiological parameter often linked to endothelial function, and provides a signal of atherosclerosis progression and thrombosis risk~\cite{zhou2023wall,katritsis2007wall}. On the other hand, the time-dependent thixotropic behaviour of blood suggests that the viscosity during the acceleration stage is different from the deceleration stage, even at the same instantaneous shear rate in pulsatile arterial flow. This causes hysteresis in flow-pressure relationships, influences energy dissipation, and ultimately complicates predictions of oscillatory shear index (OSI), another key physiological parameter that often marks the vascular disease~\cite{peiffer2013does,dintenfass1962thixotropy,vent2014blood}. Stress does not respond instantaneously to deformation due to the viscoelastic properties of blood. This leads to a phase lag between flow rate and stress, which modifies pulsatile wave propagation and can eventually shift the timing and magnitude of peak WSS in arteries~\cite{hell1989importance}. Furthermore, the viscoplasticity of blood suggests that it has a yield stress. In regions where stresses fall below this yield stress, such as stagnation zones in ventricular assist devices or post-stenotic expansions, flow can become highly nonuniform or nearly arrested. This can promote platelet activation and thrombus formation, particularly in the presence of plasma proteins such as fibrinogen, which facilitate aggregation and clotting~\cite{rampling2019red,picart1998human}. 

On the other hand, studying hemodynamics in arteries and blood-contacting medical devices is particularly important, as they directly govern physiological function and clinical performance~\cite{ku1997blood,thomas2016blood,hong2020evaluating,scafa2021cardiovascular}. It provides critical knowledge on abnormal flow patterns, which are directly linked to the onset and progression of specific cardiovascular diseases. Understanding how blood flow behaviour changes with pulsatility and how the elastic, flexible arterial wall interacts with the flow is vital for quantifying WSS, which regulates endothelial function and provides information on the progression of diseases such as atherosclerosis and thrombosis~\cite{schelbert2010anatomy,carvalho2021blood}. In arterial flows, it is essential to identify regions of low and oscillatory wall shear stress, which promote endothelial dysfunction. This is an important precursor to atherosclerosis that leads to plaque formation, arterial narrowing, and potential rupture, ultimately leading to stroke. Regions of disturbed flow and high shear gradients are also important to identify, which promote platelet activation and blood clot formation, ultimately increasing the risk of thrombosis and embolism. Furthermore, in an aneurysm, a pathologic, localised dilation (bulge) in the wall of an artery, altered hemodynamics may weaken the arterial wall and increase the risk of rupture~\cite{chalouhi2013review,tulamo2018inflammatory,syed1997coronary}. Similarly, in a stenosed artery, the narrowing accelerates flow, increasing local shear stresses that may ultimately damage the vascular lining. This information can only be obtained after a detailed hemodynamic study~\cite{berger2000flows,duraiswamy2007stented}. On the other hand, in blood-contacting medical devices, such as prosthetic heart valves and stents, a non-physiological flow condition can induce hemolysis and device-associated thrombosis~\cite{scafa2021cardiovascular,yoganathan2005flow,yoganathan2004fluid,timms2011review}. Therefore, understanding hemodynamics is essential not only for elucidating disease mechanisms but also for designing safer and more efficient medical devices, improving diagnostics, and enabling targeted therapeutic interventions.

From the aforementioned discussion, it is clear that understanding hemodynamics in arterial and medical devices is crucial for better disease diagnosis and for assessing and improving clinical performance by accounting for the complex, nonlinear rheological behaviours of blood. Keeping these in mind, several studies comprising experiments, theory, and simulations have been conducted in the literature in the past several decades on these topics. This review provides a comprehensive, up-to-date overview of those studies and their major findings on how blood rheological behaviour influences hemodynamics. Not only that, but an in-depth discussion of future scopes from a blood rheological perspective for investigating hemodynamics in arteries and blood-contacting medical devices is also provided. Ultimately, this review aims to answer whether blood's nonlinear rheological properties really matter and, if so, which are essential to consider and which are dominant for hemodynamics in arterial and medical devices, or whether it is sufficient to treat blood as a simple Newtonian fluid with constant viscosity. The review is organised as follows: first, a comprehensive review of blood rheological properties is presented in section~\ref{section2}; thereafter, section~\ref{section3} describes well-known non-Newtonian constitutive models of blood that are frequently utilised to study hemodynamics. Studies involving arterial and cardiovascular device flows, such as prosthetic mechanical heart valves and stents, considering various non-Newtonian blood properties are reviewed in sections~\ref{section3} and~\ref{section4}, respectively. Finally, the discussion and future perspectives are presented in section~\ref{section5}.           

\section{\label{section2}Review of blood rheological properties}
\begin{figure}
    \centering
    \includegraphics[width=14cm]{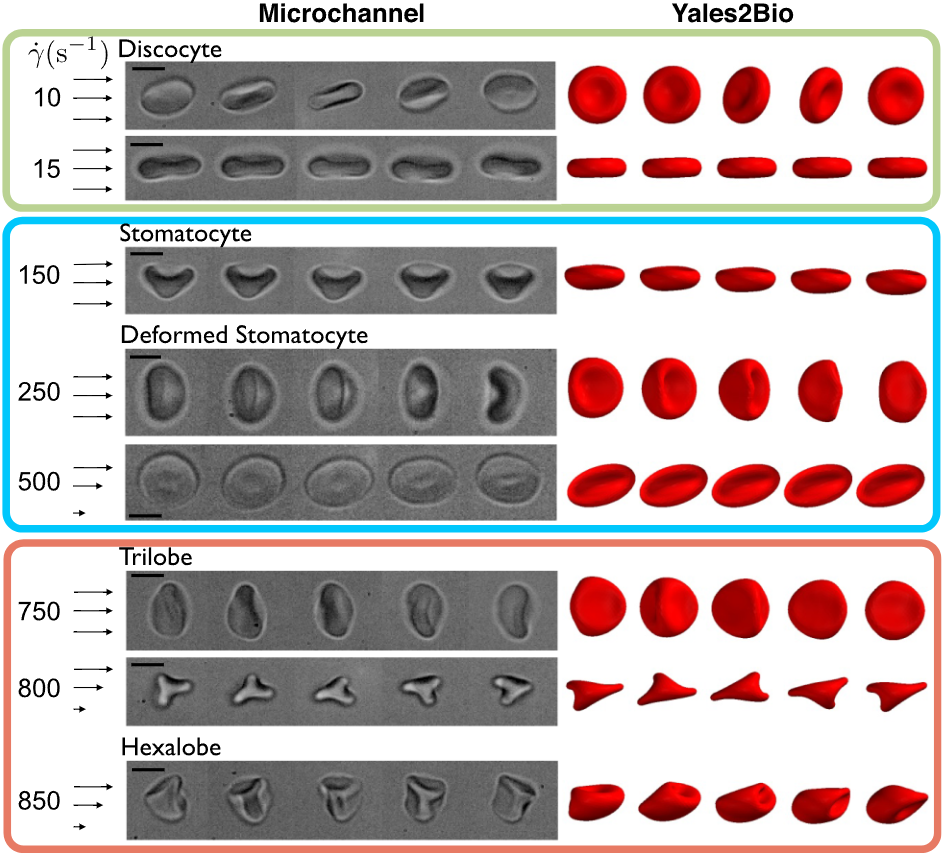}
    \caption{Visual observation of the transition of RBC shapes flowing in a microchannel with increasing shear rates $\dot{\gamma}$ obtained by Lanotte et al~\cite{lanotte2016red}. The left-hand side shows the microfluidic experimental results, whereas the right-hand side shows the corresponding simulation results obtained with their in-house software, named Yales2Bio.}
    \label{Figure1}
\end{figure}

Among the various nonlinear rheological behaviours of blood, shear thinning was probably first recognised, meaning that blood viscosity decreases with increasing deformation rate (for instance, shear rate in shear flow). Almost 100 years ago, Fahraeus and Lindqvist~\cite{fahraeus1931viscosity} conducted a groundbreaking experiment in which they measured the viscosity of blood by flowing it through capillaries of different diameters using the well-known Poiseuille equation. They found that blood viscosity decreases as capillary diameter decreases and hence concluded that it is not a constant quantity but depends on capillary diameter. This phenomenon is particularly known as the Fahraeus-Lindqvist effect~\cite{biswas2002blood}. Later, many studies have examined the mechanism of this phenomenon and found that it arises from the migration of red blood cells from the capillary wall toward the centerline under shear flow due to the generation of hydrodynamic lift forces resulting from asymmetric shear rate experienced by them, which is the highest near the wall and the lowest in the centre. This leads to the formation of a cell-free plasma layer near the wall. Since plasma has a much lower viscosity than whole blood, this near-wall lubricating layer significantly reduces viscous resistance and pressure drop. As a result, blood flows more easily in small vessels despite the increased confinement~\cite{reinke1987blood,katiukhin2014mechanism,chebbi2015dynamics}. Not only the movement of red blood cells, but their changes in shape and morphology also contribute to the shear-thinning behaviour. For instance, Lanotte et al.~\cite{lanotte2016red} experimentally and numerically showed that the shear-thinning behaviour of blood arises from shear-rate-dependent transitions in red blood cell morphology during microcirculatory flow. At low shear rates, RBCs largely retain their native discocyte shape, while increasing shear causes transitions to stomatocyte forms and, at even higher shear rates, to more complex trilobe and polylobed structures (see Figure~\ref{Figure1}). These dynamic shape changes alter cell deformation, orientation, and tank-treading motion, thereby reducing hydrodynamic resistance and decreasing the apparent viscosity of blood as the shear rate increases. Using combined microfluidic experiments and numerical simulations, the study established a direct link between RBC morphological dynamics and the macroscopic shear-thinning rheology of blood. On the other hand, Forsyth et al.~\cite{forsyth2011multiscale} investigated the relationship between red blood cell dynamics, ATP release, and blood shear-thinning behaviour under physiological shear conditions using combined microfluidics, rheology, and ATP measurements. They demonstrated that blood shear-thinning primarily originates from the transition of RBC motion from tumbling to tank-treading rather than from cell deformation alone. They further found that ATP release remains nearly constant below a critical shear stress of about 3 Pa, but increases significantly above this threshold due to increased cellular deformation. In addition, the study identified the Pannexin 1 hemichannel as the primary pathway for ATP release, while the cystic fibrosis transmembrane conductance regulator contributes mainly under high-stress deformation conditions.

\begin{figure}
    \centering
    \includegraphics[width=14cm]{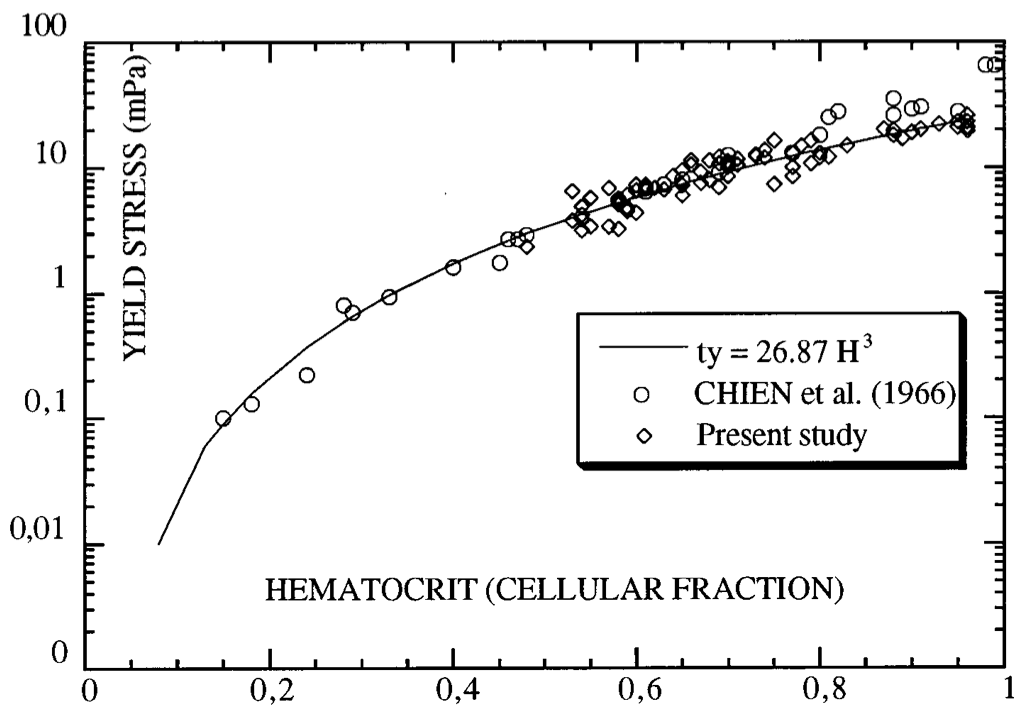}
    \caption{The variation of yield stress of human blood at different hematocrit as obtained by Picart et al.~\cite{picart1998human} along with the results of Chen et al.~\cite{chien1966effects} in the same figure at a very low shear rate of 10$^{-3}$s$^{-1}$. The solid line in this figure represents a fit of the yield stress with a cubic to the hematocrit.}
    \label{Yield}
\end{figure}

In addition to shear-thinning, the yield-stress behaviour of blood is widely reported. The early studies of Merrill et al.~\cite{merrill1963rheology} experimentally demonstrated that blood exhibits an apparent yield stress behaviour at very low shear rates due to the formation of rouleaux-like red blood cell aggregates that create a weak interconnected microstructure. They further observed that this apparent yield stress increases with hematocrit concentration but decreases with increasing temperature. Subsequent study by Chien et al~\cite{chien1966effects} showed that blood rheology at low shear rates is strongly influenced by hematocrit concentration. They found that increasing hematocrit concentration substantially increases blood viscosity and the apparent yield stress, as also observed by Merrill et al.~\cite{merrill1963rheology}, due to stronger red blood cell aggregation and cell-cell interactions. Moreover, their investigation showed that plasma proteins, particularly fibrinogen and globulins, promote rouleaux formation, leading to pronounced shear-thinning and yield-stress non-Newtonian behaviours under near-stagnant flow conditions. Their subsequent experimental study further confirmed that erythrocyte (red blood cell) aggregation is a major factor governing the non-Newtonian behaviours of blood, especially at low shear rates~\cite{chien1967blood}. Merrill et al.~\cite{merrill1966blood} showed that blood does not exhibit yield stress behaviour without fibrinogen, and therefore, they concluded that red blood cell aggregation caused by fibrinogen is the main reason for blood yield stress behaviour, which again increases with fibrinogen concentration. Chien et al.~\cite{chien1970shear} also found the same observation as that of Merrill et al.~\cite{merrill1966blood}, not only for human erythrocyte suspensions but also for canine and elephant erythrocyte suspensions. Picart et al.~\cite{picart1998human} also performed an experimental study and obtained the yield stress value of human blood at different hematocrit and found that it varies with cubic of the hematocrit, as shown in Figure~\ref{Yield}.  

\begin{figure}
    \centering
    \includegraphics[width=14cm]{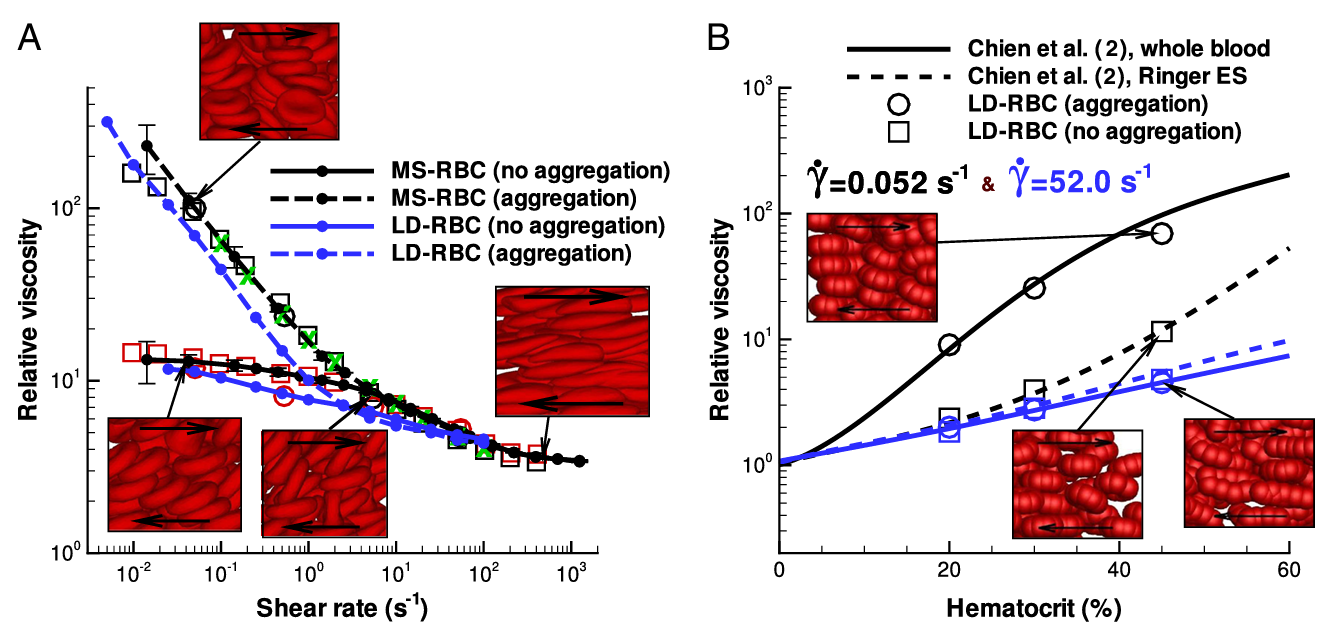}
    \caption{Dependence of blood relative viscosity on shear rate at a particular hematocrit of 45\% (a) and hematocrit at two shear rate values, namely, $\dot{\gamma} = 0.052\,s^{-1}$ and $52\,s^{-1}$ (b), as obtained in the simulation studies of Fedosov et al~\cite{fedosov2011predicting}. Both simulations are carried out at 37$^0$C. The simulation results are also compared with the corresponding prior experimental results for whole blood from Merril et al.~\cite{merrill1963rheology} (black circles) and Chien et al.~\cite{chien1966effects} (black squares). Simulations are carried out using two models: the multi-scale RBC (MS-RBC) and the low-dimensional RBC (LD-RBC), with both aggregation and non-aggregation mechanisms. The details can be found in Fedosov et al.~\cite{fedosov2011predicting}.}
    \label{Figure2}
\end{figure}

Apart from experimental investigations, particle-scale numerical simulations have also been conducted to demonstrate the non-Newtonian behaviour of blood, providing further insights into the mechanisms of these nonlinear rheological behaviours. For instance, Fedosov et al.~\cite{fedosov2011predicting} developed a coarse-grained dissipative particle dynamics (DPD) simulation framework to quantitatively predict the shear-rate and hematocrit-dependent viscosity of human blood by explicitly modelling red blood cell interactions. Their simulations revealed that reversible rouleaux aggregation at low shear rates causes a dramatic increase in blood viscosity and gives rise to apparent yield stress behaviour. The work further connected blood’s non-Newtonian rheology to RBC deformation, transient folded conformations, and complex cell dynamics occurring at intermediate shear rates, as shown in Figure~\ref{Figure2}. In addition, the study provided one of the first quantitative estimates of adhesive forces between RBCs and demonstrated the potential of such cell-resolved models for predicting rheological abnormalities associated with diseases such as malaria and diabetes. A good agreement between their numerical predictions and the corresponding experimental results for the variation of the relative viscosity with shear rate and hematocrit is evident from Figure~\ref{Figure2}. Sui et al.~\cite{sui2008dynamic} developed a three-dimensional numerical framework based on the lattice Boltzmann method (LBM) to investigate the dynamics of a single red blood cell in simple shear flow by modelling it as a deformable biconcave capsule with elastic membrane. Their simulations revealed that at high shear rates, RBC exhibits a swinging motion characterised by oscillatory inclination and membrane tank-treading, whereas decreasing shear rate triggers a transition to tumbling motion. The work successfully reproduced experimentally observed RBC dynamics and demonstrated that the suspension's apparent viscosity increases monotonically as cell motion transitions from swinging to tumbling. These findings established a direct relationship between RBC dynamical modes and the rheological behaviour of blood. Very recently, Lah et al~\cite{lah2026open} introduced the first open-boundary molecular dynamics (OBMD) framework for simulating red blood cell suspensions with explicit mass, momentum, and energy exchange across system boundaries. Using dissipative particle dynamics and a coarse-grained RBC membrane model, the simulations successfully reproduced key hemorheological features, including shear-thinning behaviour and hematocrit-dependent viscosity. The work also developed an efficient membrane insertion algorithm capable of handling high hematocrit conditions. These results established OBMD as a powerful approach for investigating realistic non-equilibrium blood flow phenomena, including pressure-driven flows, ultrasound-blood interactions, and cell-free layer formation. Perazzo et al.~\cite{perazzo2022effect} investigated how rigidified red blood cells, characteristic of sickle cell anaemia (SCA), affect blood viscosity through combined experimental and numerical simulations using the DPD method. Their results showed that even a small fraction (~5\%) of stiff cells significantly increases blood viscosity and produces a distinct rheological signature that can identify abnormal cells. The study further demonstrated that the increased viscosity arises primarily from cell rigidity rather than from cell shape or aggregation effects. These findings provide important insights into the rheological consequences of SCA and blood transfusion therapies. Readers are suggested to go through the review article by Fedosov et al~\cite{fedosov2014computational} and the textbook by Kr{\"u}ger~\cite{kruger2012computer} to know more about various computational methodologies for blood simulation and their efficiency in predicting the corresponding experimental results.  

As mentioned earlier, human blood also exhibits viscoelastic behaviour, in addition to these deformation-rate-dependent rheological characteristics such as shear-thinning and viscoplasticity. Thurston~\cite{thurston1972viscoelasticity} performed oscillatory flow experiments of blood in circular tubes to demonstrate that blood possesses significant elastic properties along with viscous properties. He conducted the experiments at a fixed frequency of 10 Hz while varying the velocity-amplitude gradient and hematocrit from 0 to 100\%. At low shear rates (less than 2 sec$^{-1}$), both viscous and elastic stresses varied linearly with velocity gradient, while increasing hematocrit strongly increased blood elasticity due to red blood cell interactions. At higher shear rates, nonlinear behaviour developed, with the viscous response approaching steady-flow viscosity and the elastic stress saturating due to red blood cell deformability and yield-stress-like effects. In a later study, Thurston and Henderson~\cite{thurston2006effects} demonstrated that the blood viscoelasticity also depends on the geometry used to measure it by performing the experiments in different geometries, such as a large cylindrical tube, a small tube, and a porous medium, which mimic a large vessel, a small vessel, and vessels with many branches and bifurcations, respectively. They found that at smaller geometries, such as in a microtube and porous medium, the viscoelasticity shows dilatancy due to increased blood cell aggregation and hardened cells. Another study by Thurston~\cite{thurston1973frequency} demonstrated that the viscoelastic response of blood strongly depends on both oscillation frequency and shear rate due to RBC aggregation and deformation. At low shear rates, blood exhibited noticeable elastic behaviour and higher viscosity, whereas increasing shear rates reduced elasticity and pronounced shear-thinning behaviour due to the disruption of cell aggregates. This was further confirmed in their study by Chien et al.~\cite{chien1975viscoelastic}, who further demonstrated that both viscosity and elasticity increase with hematocrit and are strongly influenced by cell interactions and plasma composition. Copley et al.~\cite{copley1975microscopic} revealed that the viscoelastic behaviour of blood is closely linked to the microscopic dynamics of red blood cell aggregation, disaggregation, and deformation under flow, as also proposed by others, through direct microscopic visualisation during steady and oscillatory shear experiments. The study showed that rouleaux formation and cellular structural rearrangements contribute significantly to the generation of elastic stress and time-dependent rheological behaviour. Puig-de-Morales-Marinkovic et al.~\cite{puig2007viscoelasticity} reported the first direct measurements of the complex viscoelastic modulus of isolated red blood cells using oscillatory magnetic forcing on membrane-attached ferrimagnetic microbeads. The results showed that the elastic modulus remained nearly independent of frequency and dominated the RBC response over most of the tested frequency range, whereas the frictional modulus increased with frequency, following a power-law relationship. Nonlinear mechanical responses emerged at higher deformation amplitudes, suggesting complex membrane dynamics beyond simple viscoelastic models. Their findings provided new insights into RBC deformability and highlighted the importance of this parameter in understanding microcirculatory disorders such as sickle cell disease and malaria.

\begin{figure}
    \centering
    \includegraphics[width=17cm]{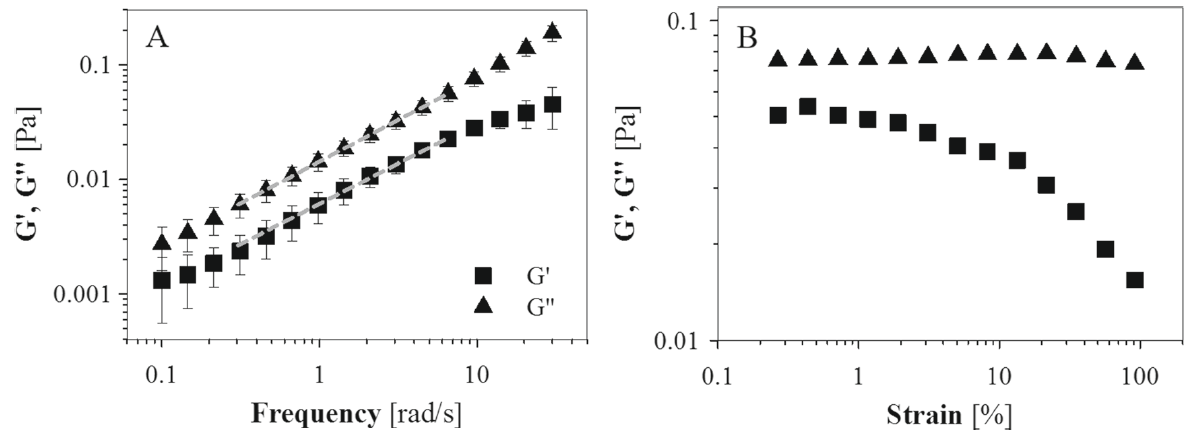}
    \caption{ Storage (G') and loss moduli (G") of whole human blood as a function of frequency (a) and as a function of strain rate (b) experimentally measured by Tomaiuolo et al.~\cite{tomaiuolo2016blood}.}
    \label{Figure3}
\end{figure}

A detailed linear viscoelasticity of blood under small deformation was experimentally measured and analysed by Tomaiuolo et al.~\cite{tomaiuolo2016blood} with the help of the small amplitude oscillatory shear (sAOS) test as a function of RBC volume fraction and aggregation. They measured storage and loss moduli over a wide range of frequencies (0.1 - 30 rad/s) of whole human blood and showed that the loss (G') modulus always predominates over the storage modulus (G"), as shown both in frequency and amplitude sweep tests presented in sub-Figures~\ref{Figure3}(a) and (b), respectively. The authors pointed out that the absence of crossover in the results of G' and G" suggests that no major change in the RBC network does not happen within the conditions encompassed in their study. They highlighted that one would expect elastic behaviour of blood to be prominent at very low deformations, where the three-dimensional RBC network structure would persist. However, due to their instrumental operational limitation, they were unable to perform experiments at really low deformation. Alves et al.~\cite{alves2013study} also conducted a similar investigation to that of Tomaiuolo et al.; however, they concluded that blood rheological measurements are strongly influenced by the measuring geometry and test duration, mainly due to secondary flows, RBC aggregation, and sedimentation. They found a crossover frequency between G' and G" in their frequency sweep experiments, unlike Tomaiuolo et al. Furthermore, they observed that the values of G' and G" initially decreased and then increased in their time sweep experiments conducted at fixed values of frequency and strain amplitude. A crossover was also seen between the two after some time, and then G' increased more than G" with time, suggesting the formation of a network structure,  a characteristic of solid structure, which increases due to an increase in network connections. Campo-Dea{\~n}o et al.~\cite{campo2013viscoelasticity} performed steady shear rheology, extensional rheology based on the capillary break-up rheometry, and passive microrheology to quantify blood's shear and extensional rheological behaviours. Their results also demonstrated that G" is larger than G', indicating liquid-like behaviour over the entire frequency range studied. Both storage and loss moduli increased with frequency, as also found by Tomaiuolo et al., though the increase was greater for the elastic component, confirming a non-negligible elastic contribution to the bulk rheology of whole human blood. 

Human blood also exhibits a more complex thixotropic rheological behaviour, i.e., time-dependent shear-thinning behaviour. It is again mainly due to the reversible aggregation and disaggregation of red blood cells. At low shear rates or during rest, RBCs form rouleaux and larger aggregates due to attractive interactions mediated by plasma proteins, such as fibrinogen, which increase viscosity. When shear is applied, these aggregates gradually break down, and the RBCs deform and align with the flow direction, thereby decreasing viscosity over time. The continuous competition between structural buildup and shear-induced breakdown of RBC networks is therefore responsible for the time-dependent rheological behaviour of blood. The same RBC aggregation and disaggregation are also responsible for blood's shear-thinning and viscoelastic behaviours, as mentioned earlier. Therefore, it should be emphasised that human blood exhibits a spectrum of rheological behaviours rather than just a particular one. Thurston~\cite{thurston1979rheological} demonstrated significant thixotropic behaviour of human blood along with viscoelasticity and shear-thinning behaviours under shear deformation, as blood viscosity was found to gradually decrease with time due to progressive breakdown of RBC microstructure and aggregates. Furthermore, in this study, he proposed a generalised Maxwell model with shear-rate-dependent parameters, developed to quantitatively analyse the steady and oscillatory flow behaviours of human blood. Stoltz and Lucius~\cite{stoltz1981viscoelasticity} also proposed that classical viscoelastic models developed for polymeric fluids, such as Maxwell-type models, could be used to explain the elastic and time-dependent response of blood. However, they emphasised that blood is more complex than ordinary polymer solutions because its rheology is governed by reversible aggregation and deformation of red blood cells rather than molecular chain entanglement alone. In line with Stoltz and Lucius, Huang et al.~\cite{huang1995viscoelastic} actually proposed an equation based on the generalised Maxwell model by incorporating the thixotropic component into the viscous component of the relaxation modulus and obtained a good representation of the corresponding experimental hysteresis loops, pertaining to thixotropic behaviour, for both normal and diabetic blood. Quemada and Droz~\cite{quemada1983blood} proposed a structural Maxwell-type viscoelastic model for blood that incorporates time-dependent viscosity and elasticity through a structure parameter representing the evolution of rouleaux and RBC networks. Their model successfully explained stress relaxation, stress growth, and the characteristic stress-overshoot behaviour observed in the corresponding experiments under constant shear. 

So far, the shear rheological properties of human blood have been discussed. However, human blood also exhibits quite distinct rheological behaviour, undergoing extensional flows, unlike a simple Newtonian fluid. For instance, Sousa et al.~\cite{Sousa2018} investigated the uniaxial extensional rheological response of whole human blood using capillary breakup (CaBER) experiments. In this experiment, a small volume of fluid, for instance, blood in this case, is placed between two circular plates, which are rapidly separated to form a liquid filament. The evolution of the minimum filament diameter $D(t)$ is recorded using a high-speed camera; the filament-thinning dynamics are then used to characterise extensional viscosity and relaxation time. By performing this experiment, they found that human blood exhibits finite extensional relaxation times ranging from 114 $\pm$ 30 to 259 $\pm$ 47 $\mu$s in air and oil, respectively. Furthermore, they observed that an increase in RBC concentration delays filament thinning and increases the time to break up. Recent studies have found that even human blood plasma exhibits considerable viscoelastic behaviour under extensional flow. For instance, Brust et al.~\cite{brust2013rheology} performed both shear and extensional flow experiments with human blood plasma. In shear flows, plasma showed Newtonian behaviour; however, in extensional flows with a micro-contraction-and-expansion geometry, it exhibited greater resistance to flow than pure water, which they attributed to its viscoelasticity. A subsequent simulation study by Varchinis et al.~\cite{Varchanis2018} further confirmed the presence of viscoelasticity of blood plasma.

\section{\label{section3}Summary of non-Newtonian constitutive models for hemodynamics}
This section discusses the constitutive models that are used in theoretical or numerical analyses to study hemodynamics in various arteries and to design and evaluate the performance of blood-contacting medical devices. The parameters of these constitutive models are often obtained by fitting experimental data on human blood under different rheological flows, such as steady shear or small-amplitude oscillatory shear. Only the widely used non-Newtonian constitutive models are summarised here, which are broadly classified into three categories, namely, inelastic generalised Newtonian fluid models, viscoelastic models, and time-dependent models. In the case of inelastic models, the apparent viscosity is a function of the rate of deformation only, whereas for viscoelastic and time-dependent models, it is additionally a function of deformation history and a structure parameter, respectively.  
\subsection{Inelastic generalised Newtonian fluid models} 
\begin{itemize}
    \item \textbf{Power-law or Ostwald-de Waele model:} The power-law constitutive model was proposed by French scientist Wilhelm Ostwald, a Nobel prize-winning chemist, and Armand de Waele, a British chemist, almost at the same time in the early 20th century~\cite{ostwald1925ueber,waele1923viscometry}. This is the reason why this model is also called the Ostwald-de Waele model. According to this constitutive model, the apparent viscosity of blood can be evaluated as 
    \begin{equation}
       \eta = m \left( \dot{\gamma} \right)^{n-1}
       \label{eq:powerlaw}
    \end{equation}
    In the above equation, $m$ and $n$ are known as the flow consistency index (with SI unit of Pa $\cdot$ s$^{n}$) and flow behaviour index (dimensionless), respectively. The main advantage of this constitutive model is its simple mathematical form for capturing shear-rate-dependent viscosity. Due to its simple form, it is comparatively easy to tackle both in theoretical analysis and computer simulations. The model has only two parameters, namely, $m$ and $n$, and therefore, it is relatively easy to fit the experimental data and predict the flow curve. Despite these advantages, this model also has some significant limitations, such as it can not predict the limiting values of the apparent viscosity exhibited by a fluid at very high (infinite-shear rate viscosity, $\eta_{\infty}$) or low shear (zero-shear rate viscosity, $\eta_{0}$) rates.
    
  \item \textbf{Carreau-Yasuda model:} In 1972, Pierre J. Carreau, a renowned rheologist, proposed a model that can predict zero-shear and infinite-shear viscosities in the limit of low and high shear rates, respectively. The model named after him, the Carreau model, has the following expression
\begin{equation}
    \eta = \eta_{\infty} + \left(\eta_{0} - \eta_{\infty}\right) \left[1 + \left( \lambda \dot{\gamma} \right)^{2}\right]^{\frac{(n-1)}{2}}
    \label{Carreau}
\end{equation}
In the above expression, $\lambda$ is a time constant (relaxation time with the unit of $s$), determining the onset of shear-thinning or shear-thickening behaviours in the apparent viscosity vs shear rate curve. $n$ is the same power-law index as there for the power-law fluid model. In the limit of $\dot{\gamma} \rightarrow 0$, $\eta \rightarrow \eta_{0}$ as the term in the square bracket tends to 1. On the other hand, in the limit of $\dot{\gamma} \rightarrow \infty$, the term in the square bracket tends to 0 for a shear-thinning fluid as the exponent $(n - 1)$ is negative, and as a result, $\eta \rightarrow \eta_{\infty}$. For a Newtonian fluid (i.e., $n = 1$), $\eta = \eta_{0} = \mu$. Note that this model is particularly applicable for shear-thinning fluids, i.e., $n < 1$. Later, in 1981, K. Yasuda~\cite{yasuda1981shear} modified the Carreau model by introducing an additional fitting parameter, yielding the following form.
\begin{equation}
    \eta = \eta_{\infty} + \left(\eta_{0} - \eta_{\infty}\right) \left[1 + \left( \lambda \dot{\gamma} \right)^{a}\right]^{\frac{(n-1)}{a}}
    \label{CarreauYasuda}
\end{equation}
Here, the additional fitting parameter $a$ (dimensionless) is the transitional parameter that adjusts the width of the transition region between the Newtonian plateau and the power-law regions. Notably, $a = 2$ for the original Carreau model. After this modification by Yasuda, the Carreau model is better known as the Carreau-Yasuda model. This model is also widely used for blood flow analysis. 

\item \textbf{Walburn-Schneck model:} This is an empirical rheological model developed by F. J. Walburn and D. J. Schneck in the late 1970s specifically to describe the apparent viscosity of human blood as a function of shear rate, hematocrit, and plasma protein concentration~\cite{walburn1976constitutive}. Unlike other models, this model explicitly incorporates physiological parameters, making it particularly useful for patient-specific blood rheology. According to this model, the apparent viscosity of blood can be evaluated as
\begin{equation}
    \eta = C_{1}e^{(C_{2}H)}e^{(C_{4} \text{TPMA}/H^{2})}\dot{\gamma}^{-C_{3}H}
\end{equation}
Where $H$ is the hematocrit, TPMA is the total plasma protein concentration, $\dot{\gamma}$ is the shear rate, and $C_{1}-C_{4}$ are the empirical constants determined by fitting with the experimental data. 

\item \textbf{Herschel-Bulkley model:} This model was introduced by Winslow Herschel and Ronald Bulkley in 1926, which can account for the yield stress of blood along with shear-thinning behaviour~\cite{herschel1926konsistenzmessungen}. As per this model, the apparent viscosity of blood can be calculated as follows.
\begin{equation}
    \eta = \frac{\tau_0}{\dot{\gamma}} + K \dot{\gamma}^{n-1}
\end{equation}
Here, $K$ is the flow consistency index (with SI unit of Pa $\cdot$ s$^{n}$), $n$ is the flow behaviour index (dimensionless), and $\tau_0$ is the yield stress (with SI unit of $Pa$). Note that when blood does not exhibit yield stress, i.e., $\tau_0 = 0$, the model reduces to that of the power-law model. This is a three-parameter $(\tau_0, K, n)$ constitutive model. Furthermore, in the limit of $n = 1$, the Herschel-Bulkley model reduces to the Bingham model (sometimes called the Bingham plastic model), which is also used in blood flow analysis when only the yield stress needs to be incorporated. This model was originally introduced to describe the flow of mud or slurries exhibiting finite yield stresses~\cite{bingham1917investigation}. One of the biggest drawbacks of these models is that the apparent viscosity tends to infinity when $\dot{\gamma} \rightarrow0$, as can be seen from the above equation. Furthermore, as per this constitutive equation, the shear rate should be zero when the shear stress is less than the yield stress, i.e., $\dot{\gamma = 0}$ when $\tau < \tau_0$. This discontinuity is difficult to handle in numerical simulations, leading to convergence issues. Therefore, one has to use regularisation techniques to address this inherent problem in the constitutive model, such as Papanastasiou regularisation~\cite{Papanastasiou1987}. According to this regularisation technique, the apparent viscosity is calculated as, for instance, for the Herschel-Bulkley model, 
\begin{equation}
    \eta = K \dot{\gamma}^{n-1} + \frac{\tau_0}{\dot{\gamma}}\left(1-e^{-m\dot{\gamma}}  \right)
\end{equation}
Where $m$ is the regularisation parameter (with unit of $s$). As $m \rightarrow \infty$, this expression leads to the ideal Herschel-Bulkley model while remaining finite at very low shear rates.

\item \textbf{Casson model:} This model was originally introduced by Norman Casson to describe the flow dynamics of concentrated pigment suspensions, such as printing inks, which exhibit both yield stress and shear-thinning behaviours~\cite{casson1959flow}. Subsequently, the model has also become popular for studying hemodynamics, as blood exhibits these rheological behaviours. As per this constitutive model, the apparent viscosity of blood can be calculated as follows:
\begin{equation}
    \eta = \eta_c + 2 \sqrt{\frac{\eta_c \tau_0}{\dot{\gamma}}} + \frac{\tau_0}{\dot{\gamma}}
\end{equation}
Where $\eta_c$ is the blood plastic viscosity (with unit of $Pa.s$) and $\tau_0$ is again the yield stress. Likewise, the Herschel-Bulkley or Bingham plastic model, and the Casson model, also have the drawback of a discontinuity in the prediction of apparent viscosity as $\dot{\gamma} \rightarrow 0$; therefore, one has to consider a regularisation technique to get rid of this problem. 

\item \textbf{Quemada model:} This constitutive model was originally developed to describe the rheological behaviours of concentrated colloidal suspensions by Diego Quemada in the late 1970s and early 1980s~\cite{quemada1977rheology}. Unlike purely phenomenological models, such as power-law or Casson models, this model incorporates the influence of red blood cell concentration (hematocrit) on the apparent viscosity variant. Therefore, it is considered one of the earliest microstructure-inspired semi-empirical rheological models for blood. As per this model, the apparent viscosity is evaluated as follows.
\begin{equation}
    \eta = \eta_p \left(1 -  \frac{1}{2}k(\dot{\gamma}) H\right)^{-2}
\end{equation}
Where $\eta_p$ is the plasma viscosity, $H$ is the hematocrit, and $k(\dot{\gamma})$ is the intrinsic viscosity parameter, which can be calculated as below
\begin{equation}
    k(\dot{\gamma}) = \frac{k_0 + k_\infty \sqrt{\dot{\gamma}/\dot{\gamma}_c}}{1+\sqrt{\dot{\gamma}/\dot{\gamma}_c}}
\end{equation}
Where $k_0$, $k_\infty$, and $\dot{\gamma}_c$ are the interaction coefficient at zero shear, at infinite shear, and characteristic shear rate, respectively. 

\item \textbf{Cross model}\newline
In 1965, Malcolm M. Cross proposed a constitutive model, named after him the Cross viscosity model, which has the following form~\cite{cross1965rheology}

\begin{equation}
    \eta = \eta_{\infty} + \frac{\eta_{0} - \eta_{\infty}}{1 + \left(\lambda \dot{\gamma}\right)^{n}}
    \label{crossModi}
\end{equation}
In the above equation, $\lambda$ is positive and interpreted as the time constant having the unit of time, which has the same significance as in the Carreau-Yasuda model. Furthermore, $n$ is an arbitrary dimensionless power-law exponent that decides the slope of the apparent viscosity curve in the power-law region. In the limits of $\dot{\gamma} \rightarrow 0$ and~$\infty$, $\eta \rightarrow \eta_{0}$ and $\eta_{\infty}$, respectively, as can be seen from equation~\ref{crossModi}. The Cross viscosity model has four fitting parameters, namely, $\eta_{0}$, $\eta_{\infty}$, $\lambda$, and $n$.     
\end{itemize}

\begin{figure}
    \centering
    \includegraphics[width=12cm]{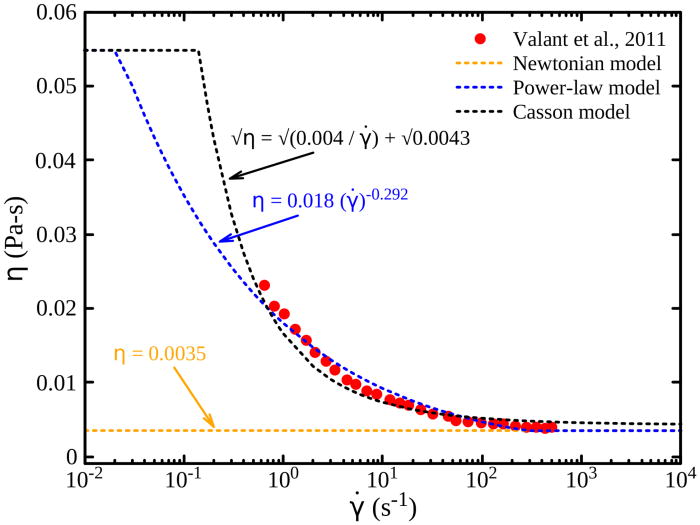}
    \caption{Fitting of the experimental blood steady shear rheology data by Vlatant et al.~\cite{ZupaniValant2011} with two inelastic constitutive models, namely, power-law and Casson models~\cite{Chauhan2024}.} 
    \label{ShearFit}
\end{figure}

Figure~\ref{ShearFit} demonstrates the fitting of the experimental data on steady shear rheology by Valant et al.~\cite{ZupaniValant2011} with two inelastic constitutive models, namely, the power-law and Casson models~\cite{Chauhan2024}. The corresponding fitting parameters for these two models are also given in the expressions shown in the figure. 

\subsection{Viscoelastic models}
\begin{itemize}
    \item \textbf{Upper Convected Maxwell (UCM) and Oldroyd-B models:} Oldroyd, in his seminal paper~\cite{oldroyd1950formulation}, proposed that the total extra stress tensor in a viscoelastic fluid like human blood is simply the sum of the two components, i.e.,  
\begin{equation}
\label{oldroydB}
    {\boldsymbol{\tau}} = {\boldsymbol{\tau}}_{s}+{\boldsymbol{\tau}}_{p} = \eta_{s} \dot{{\boldsymbol{\gamma}}} + {\boldsymbol{\tau}}_{p} 
\end{equation}
The solvent contribution (in the case of blood, it is the plasma) is calculated using Newton's law of viscosity, i.e., ${\boldsymbol{\tau}}_{s} = \eta_{s} \dot{{\boldsymbol{\gamma}}}$ where $\eta_s$ is the solvent viscosity, whereas the fluid microstructure contribution (for example RBC in the case of human blood) is evaluated using the Upper Convected Maxwell model as follows.
\begin{equation}
\label{ucm}
    {\boldsymbol{\tau_p}} + \lambda \overset{\nabla}{{\boldsymbol{\tau_p}}} = \eta \,\dot{{\boldsymbol{\gamma}}}
\end{equation}
Where $\overset{\nabla}{{\boldsymbol{\tau_p}}}$ is the upper convected derivative of the stress tensor given by the expression as $\frac{\partial {\boldsymbol{\tau_p}}}{\partial t} + \boldsymbol{u}\cdot\boldsymbol{\nabla{\boldsymbol{\tau_p}}} - \left( \boldsymbol{\nabla} \boldsymbol{u}\right)^{T}\cdot {\boldsymbol{\tau_p}} - {\boldsymbol{\tau_p}}\cdot\left( \boldsymbol{\nabla} \boldsymbol{u} \right)$. Sometimes, the Oldroyd-B model is also represented in another form in terms of the total extra stress tensor, given below
\begin{equation}
    {\boldsymbol{\tau}} + \lambda_{1}\overset{\nabla}{{\boldsymbol{\tau}}} = \eta \left( \dot{{\boldsymbol{\gamma}}} + \lambda_{2} \overset{\nabla}{\dot{{\boldsymbol{\gamma}}}}\right)
\end{equation}
Where $\lambda_{1}$ is the fluid relaxation time, $\lambda_{2} \left( =\frac{\eta_{s}}{\eta_{s}+\eta_{p}} \lambda_{1}\right)$ is the retardation time, $\eta\, (=\eta_{s} + \eta_{p})$ is the total zero-shear viscosity, and $\overset{\nabla}{\dot{{\boldsymbol{\gamma}}}}$ is the upper convected derivative of the strain-rate tensor, equal to $\overset{\nabla}{\dot{{\boldsymbol{\gamma}}}} = \frac{\partial \dot{{\boldsymbol{\gamma}}}}{\partial t} + \boldsymbol{u} \cdot \boldsymbol{\nabla} \dot{{\boldsymbol{\gamma}}} - \left( \boldsymbol{\nabla} \boldsymbol{u}\right)^{T} \cdot \dot{{\boldsymbol{\gamma}}} - \dot{{\boldsymbol{\gamma}}} \cdot \boldsymbol{\nabla} \boldsymbol{u}$. After taking the upper convected derivative of Eq.~\ref{oldroydB} and then multiplying by $\lambda$ on both sides, and then substituting the expression for $\lambda \overset{\nabla}{{\boldsymbol{\tau}}}$ from the UCM model, we can get the above equation after some rearrangement. Note that in the limit of $\eta_{s} = 0$, the above equation reduces to the UCM model, whereas it becomes the Newtonian fluid expression when $\lambda_{1} = 0$. The value of the retardation time is generally much smaller than the fluid microstructure relaxation time, which basically represents a delayed viscous response due to solvent contribution. 

\item \textbf{Phan-Thien-Tanner (PTT) model:} The PTT (Phan-Thien-Tanner) viscoelastic model, introduced by Nam Phan-
Thien and Roger I. Tanner~\cite{thien1977new}, which can depict both shear-thinning and viscoelastic properties of blood. In particular, the simplified Phan-Thien-Tanner (sPTT) model is used for blood flow simulations, which is mathematically expressed as:  
\begin{equation} \label{eq:sPTT_constitutive}
   \left( 1+ \frac{\lambda \,\epsilon \,\text{tr}({\boldsymbol{\tau_p}})}{\eta_p}   \right) {\boldsymbol{\tau_p}} + \lambda \overset{\nabla}{{\boldsymbol{\tau_p}}}=\eta_{p} \,\dot{{\boldsymbol{\gamma}}}
\end{equation}
Where $\overset{{\nabla}}{{{\boldsymbol{\tau_p}}}}$ again denotes the upper-convected derivative of ${{\boldsymbol{\tau_p}}}$ as already written for the Oldroyd-B and UCM models. Here, $\lambda$, $\epsilon$, and $\eta_p$ denote the relaxation time, extensibility coefficient, and viscosity of the viscoelastic blood component, respectively. Many times, a multi-mode version of this constitutive model is used to better fit the experimental rheological data of blood, wherein the total extra stress component is calculated by summing up the individual contributions from each mode as follows
\begin{equation} \label{eq:sPTT}
   {\boldsymbol{\tau_p}} = \sum_{k=1}^{N} {\boldsymbol{\tau_p^k}} 
\end{equation}
Where $k$ is the number of modes used in the fitting.  
\end{itemize}

\subsection{Thixo-elasto-viscoplastic models}
So far, the models described can only predict the shear-thinning, viscoplastic, and viscoelastic behaviours of blood. However, as mentioned earlier, blood also exhibits thixotropic behaviour. In fact, blood exhibits all these rheological behaviours together. Therefore, in the literature, suitable constitutive models have also been developed to predict all these rheological behaviours together. These models are collectively referred to as thixo-elasto-viscoplastic models of blood. 

\begin{itemize}
    \item \textbf{TEVP model:} The thixo-elastoviscoplastic (TEVP) model proposed by Spyridakis et al.~\cite{Spyridakis2023} describes the rheological response of human blood by accounting for its viscoelasticity, yielding, shear-thinning behaviour, and time-dependent microstructural evolution. As per this model, the total rate-of-deformation tensor is defined as
\begin{equation}
    \dot{\boldsymbol{\gamma}}
    =
    \nabla \boldsymbol{u}
    +
    \left(\nabla \boldsymbol{u}\right)^{T},
\end{equation}
Where $\boldsymbol{u}$ is the velocity vector. The total deformation rate is decomposed into elastic and viscoplastic contributions as
\begin{equation}
    \dot{\boldsymbol{\gamma}}
    =
    \dot{\boldsymbol{\gamma}}_{e}
    +
    \dot{\boldsymbol{\gamma}}_{vp}.
\end{equation}

The elastic contribution is related to the upper-convected derivative
of the extra stress tensor through
\begin{equation}
    \dot{\boldsymbol{\gamma}}_{e}
    =
    \frac{1}{G_t}
    \overset{\nabla}{\boldsymbol{\tau}},
\end{equation}
Where the structure-dependent elastic modulus is given by
\begin{equation}
    G_t
    =
    \frac{G_0}{\lambda^{n_g}}.
\end{equation}

The upper-convected derivative of the extra stress tensor is defined as
\begin{equation}
    \overset{\nabla}{\boldsymbol{\tau}}
    =
    \frac{\partial\boldsymbol{\tau}}{\partial t}
    +
    \boldsymbol{u}\cdot\nabla\boldsymbol{\tau}
    -
    (\nabla\boldsymbol{u})^{T}\cdot\boldsymbol{\tau}
    -
    \boldsymbol{\tau}\cdot\nabla\boldsymbol{u}.
\end{equation}

The viscoplastic contribution is expressed as
\begin{equation}
    \dot{\boldsymbol{\gamma}}_{vp}
    =
    \frac{
    \exp\left[
    \epsilon_{\mathrm{PTT}}
    \dfrac{\operatorname{tr}(\boldsymbol{\tau})}{G_t}
    \right]
    }{\eta_t}
    \boldsymbol{\tau},
\end{equation}
Where the structure-dependent plastic viscosity is
\begin{equation}
    \eta_t
    =
    \frac{\lambda^{m_1}}
    {1-\lambda}
    \eta_{0,\mathrm{RBC}}.
\end{equation}

Combining the elastic and viscoplastic contributions gives the
constitutive equation for the extra stress:
\begin{equation}
\frac{1}{G_0\lambda^{-n_g}}
\overset{\nabla}{\boldsymbol{\tau}}
+
\frac{1-\lambda}
{\eta_{0,\mathrm{RBC}}\lambda^{m_1}}
\exp\left[
\epsilon_{\mathrm{PTT}}
\frac{\operatorname{tr}(\boldsymbol{\tau})}{G_t}
\right]
\boldsymbol{\tau}
=
\nabla\boldsymbol{u}
+
(\nabla\boldsymbol{u})^T
\end{equation}

or, equivalently,

\begin{equation}
\frac{\lambda^{n_g}}{G_0}
\overset{\nabla}{\boldsymbol{\tau}}
+
\frac{1-\lambda}
{\eta_{0,\mathrm{RBC}}\lambda^{m_1}}
\exp\left[
\epsilon_{\mathrm{PTT}}
\frac{\operatorname{tr}(\boldsymbol{\tau})}
{G_0\lambda^{-n_g}}
\right]
\boldsymbol{\tau}
=
\dot{\boldsymbol{\gamma}}.
\end{equation}

The evolution of the structural parameter $\lambda$ is described by
\begin{equation}
\frac{d\lambda}{dt}
=
k_1(1-\lambda)
+
|\dot{\gamma}|
\exp\left[-(t_R|\dot{\gamma}|)^2\right](1-\lambda)
-
k_2
\left[
(\nabla\boldsymbol{u})^T:\boldsymbol{C}
\right]
\lambda^{m_2},
\end{equation}
Where $\lambda$ represents the degree of aggregation of the red blood cells, with $\lambda=1$ corresponding to a fully structured state and smaller values representing progressively broken-down rouleaux structures. Furthermore, $\boldsymbol{C}$ is the conformation tensor which accounts for the shape and rouleaux structure under deformation defined as $\boldsymbol{C} = \frac{\boldsymbol{\tau}}{G_t}+ \boldsymbol{I}$ where $\boldsymbol{I}$ is the identity tensor. The first term in the above equation represents Brownian restructuring, the second term accounts for shear-induced RBC aggregation, and the third term represents the destruction of rouleaux structures. The Gaussian function $\exp[(t_R|\dot{\gamma}|)^2]$ ensures that shear-induced aggregation is significant only over a limited range of shear rates. The structure-dependent model, therefore, captures both rebuilding and breakdown of the blood microstructure. The model contains a total of 9 parameters (hence, it is also called the TEVP 9 model), namely, $G_0$ is the reference elastic modulus, $n_g$ is the elasticity-structure exponent, $\epsilon_{PTT}$ is the PTT nonlinear parameter,  $\eta_{0,\text{RBC}}$ is the RBC zero-shear viscosity, $m_1$ is the viscosity structure exponentt, $k_1$ is the Brownian rebuilding rate, $t_R$ is the flow-aggregation time scale, $k_2$ is the destruction rate constant, and $m_2$ is the destruction structure exponent. This model is developed primarily by modifying the author's previous models~\cite{varchanis2019modeling,giannokostas2020advanced}.   

\end{itemize}

\section{\label{section4}Impact of blood rheology on arterial flows}
A large number of studies, including theoretical, numerical, and experimental work, have been carried out on arterial blood flow. Some excellent review articles are already present on this; for instance, see the articles by Ku~\cite{ku1997blood}, Thomas and Suman~\cite{thomas2016blood}, Carvalho et al.~\cite{carvalho2021blood}, Pandey et al.~\cite{pandey2020review}. However, most of these review articles did not address the rheological perspective of blood in arterial flows, which is the aim of this review. \\

The theoretical study by Shukla et al.~\cite{shukla1980effects} in 1980 was probably the first thorough investigation of the effects of blood rheological behaviour on hemodynamics in a stenosed artery, considering two rheological models for blood, namely, the power-law and Casson models. They found that flow resistance increases with stenosis severity; however, this increase was less pronounced when blood was treated as non-Newtonian rather than Newtonian. Not only the flow resistance, but also the wall shear stress, was found to increase less, and the velocity field was less disturbed when non-Newtonian blood rheology was considered. Moravec and Liepsch~\cite{moravec1983flow,liepsch1984pulsatile} conducted both an experimental study using a laser Doppler anemometer and a numerical study on arteries with bends and bifurcations, considering blood both as Newtonian (aqueous glycerol) and non-Newtonian (aqueous solution of polyacrylamide) fluids, under both steady and pulsatile flow conditions. They found that the impact of blood's non-Newtonian behaviour was prominent distal to the bifurcation, where the recirculation zone was much larger when blood was treated as non-Newtonian compared to Newtonian, particularly in pulsatile flows. A substantial difference in the flow hemodynamics was also found in the experimental study of Ku and Liepsch~\cite{ku1986effects}, particularly in the main tube where secondary flows and flow separation happen in a 90-degree bifurcated artery. Additionally, they observed differences in flow behaviour when assuming the artery to be rigid or elastic. Chaturani and Ponnalagar Samy~\cite{chaturani1985study} conducted a numerical study of a stenosed artery by assuming blood as a non-Newtonian Herschel-Bulkley fluid. They varied various model parameters of this constitutive model, such as the power-law index, flow consistency index, and yield stress, and found that the flow resistance and wall shear stress increased compared with the Newtonian model. This is in contradiction to that observed by Shukla et al.~\cite{shukla1980effects}. Chakravarty~\cite{chakravarty1987effects} performed a theoretical analysis considering the deformability of the stenosed tube and the power-law model of the blood with arbitrary values of the model parameters. His results showed that flow resistance increases with higher power-law exponents. In a subsequent study, Chakravarty and Datta~\cite{chakravarty1989effects} used the same geometrical configuration but employed the Herschel-Bulkley constitutive model for blood, with realistic parameters matched to blood properties. They found an overestimation of the resistance to flow through the stenosed area when assuming blood to be a simple Newtonian fluid rather than a non-Newtonian one. They extended this study to a multi-stenosed case as well, in which they found that the presence of multi-stenosis significantly influenced hemodynamics~\cite{chakravarty1990dynamic}. Nakamura and Sawada~\cite{nakamura1988numerical} performed a finite element method-based numerical study on steady laminar flow of blood, modelled using the biviscosity model, through a stenosed artery. Their study revealed that the non-Newtonian properties of blood reduce the axial force acting on the stenosis relative to the Newtonian assumption. Therefore, they suggested that the non-Newtonian blood property inhibits the post-stenotic dilation controlled by this axial force.                       
\begin{figure}
    \centering
    \includegraphics[width=17cm]{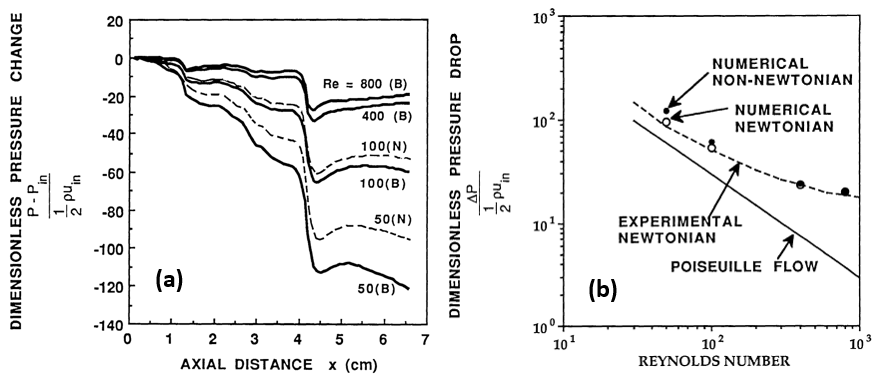}
    \caption{Pressure drop variation for flows through a stenosed artery by assuming blood as a simple Newtonian fluid and a shear-thinning power-law non-Newtonian fluid at different Reynolds numbers~\cite{cho1991effects}  (a) Variation along the length of the artery $x$ (b) Difference between inlet and outlet of the artery. Here, $u_{in}$ and $p_{in}$ are the blood inlet velocity and pressure at the inlet, whereas $\rho$ is the blood density. Furthermore, in sub-figure (a), "B" stands for the non-Newtonian model, whereas "N" refers to the Newtonian model.}
    \label{Figure4}
\end{figure}

Choi and Kensey~\cite{cho1991effects} performed a detailed numerical analysis of steady blood flow through a severely blocked $(78\%)$ stenosed artery by considering blood as a non-Newtonian power-law fluid, by evaluating the parameters of the model by fitting the experimental data of blood shear viscosity. They found that the differences in predictions of physiological parameters, such as pressure drop and wall shear stress, between Newtonian and non-Newtonian blood models depend on the Reynolds number. For instance, Figure~\ref{Figure4} shows the variation in the pressure drop along the length of the artery (a) as well as between the inlet and outlet of the artery (b). In both cases, they observed that when the Reynolds number is high, the difference between the two models is negligible. However, a noticeable difference was observed at low Reynolds numbers, with the non-Newtonian model predicting higher values (12.4\% and 26.8\% at $Re = 100$ and 50, respectively) than the Newtonian model. A similar trend was also observed in the variation of wall shear stress. Luo and Kuang~\cite{luo1992non} also found a larger pressure drop and wall shear stress for the non-Newtonian blood viscosity model than the Newtonian one for the flow through an arterial stenosis at the Reynolds number of 100. However, they also observed that the recirculation region downstream of the stenosis is smaller for the former than for the latter, suggesting that the flow is more stable when the blood's non-Newtonian properties are accounted for. In contrast, Lou and Yang~\cite{chien1973ultrastructural} in their simulation study using the Casson non-Newtonian blood model found that the non-Newtonian blood property does not have a dramatic change in the flow patterns; however, they also found an increase in flow resistance and wall shear stress due to non-Newtonian properties of the blood for the flow through the aortic bifurcation. Baaijens et al.~\cite{baaijens1993numerical} also concluded from their study on the two-dimensional carotid artery bifurcation that the non-Newtonian properties of blood modelled using the power-law and Casson models have little influence on the overall flow structure. However, some quantitative differences were observed in the pressure and wall shear stress values. For instance, they also observed 25\% and 5\% higher pressure and wall shear stress values, respectively, with generalised Newtonian fluid models than with the Newtonian model. Tandon et al.~\cite{tandon1993model} found that the non-Newtonian nature of blood, modelled using Casson's model, reduces the peak wall shear stress and the recirculation zone formed downstream of a stenosis in the case of flow through a double-stenosed artery. Das and Batra~\cite{Das1995} performed a numerical study for the flow of blood modelled by the Casson model through a stenosed artery for both rigid and permeable artery walls. They found that the flow resistance and wall shear stress decrease as the Casson number (which quantifies the extent of non-Newtonian behaviour of blood in terms of the non-dimensional yield stress) and the permeability parameter increase, and that their values were always lower when the blood was assumed to be Newtonian. Tandon and Rana~\cite{Tandon1995} simulated the flow of blood through a stenosed artery by assuming blood as a suspension of cells in plasma. They considered three distinct zones in their simulations: a cell-free layer near the artery wall, a central core region, and an annular layer of a non-Newtonian biviscous fluid. Once again, they found reduced wall shear stress and a shorter flow reversal zone, as others have also reported.

\begin{figure}
    \centering
    \includegraphics[width=17cm]{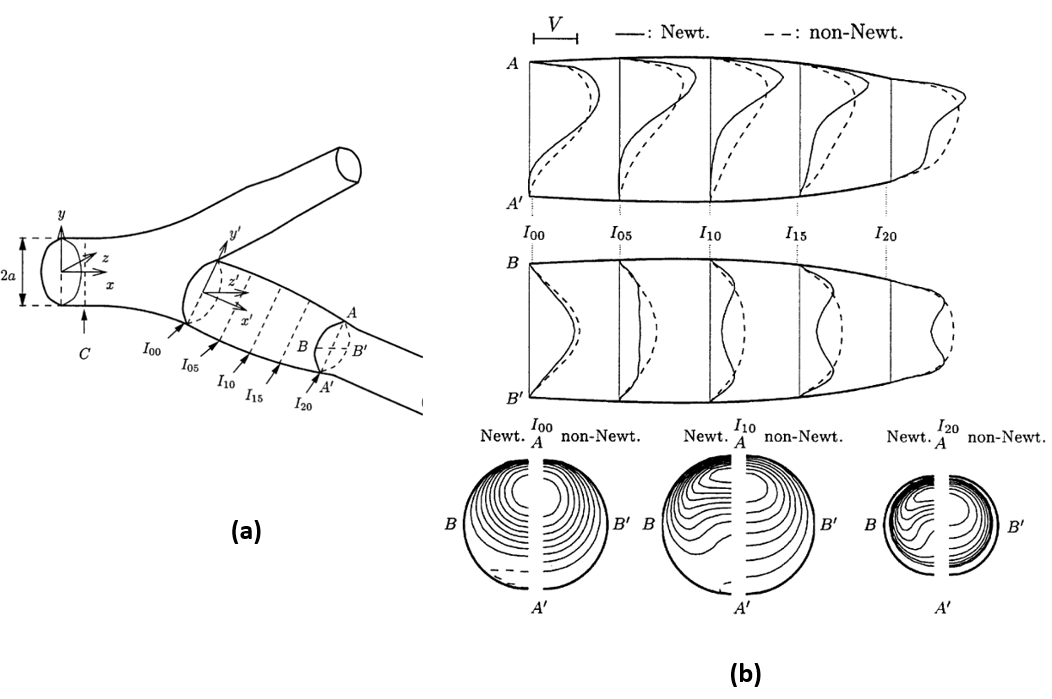}
    \caption{(a) Schematic of the carotid bifurcated geometry with the measurement locations both in experiments and numerical simulations by Gijsen et al~\cite{Gijsen1999_1}. (b) The comparison on the axial velocity profiles in the symmetry plane A-A' (first row) and perpendicular to this plane B-B' (second row) and streamline patterns (third row) between Newtonian and non-Newtonian blood models.}
    \label{Figure5}
\end{figure}

Dutta and Tarbell~\cite{Dutta1996} obtained an analytical solution for blood flow through an elastic straight artery by considering two constitutive models, namely, the shear-thinning power-law model and the generalised Maxwell model, which accounts for both shear-thinning and oscillatory flow viscoelasticity. They concluded that fluid elasticity has minimal influence on flow rate and WSS predictions under physiological conditions in large arteries; hence, the assumption of shear-thinning blood viscosity is sufficient to predict these flow behaviours. Furthermore, they reported that wall motion in elastic arteries has a far greater effect than the non-Newtonian effect on predictions of flow rate and WSS. Tu and Deville~\cite{Tu1996} performed finite element method-based numerical simulations of a rigid artery with varying degrees of stenosis, under both steady and pulsatile inflows, considering four blood rheological models, namely, Herschel-Bulkley, Bingham, power-law, and Newtonian. Their study revealed that the non-Newtonian blood properties exhibit greater disturbances in the flow profiles downstream of the stenosis than the constant-viscosity Newtonian ones, but show lower wall shear stress values. In contrast, for a carotid artery, Wells et al.~\cite{Wells1996} found slightly higher values of wall shear stress and its gradients for the non-Newtonian power-law blood model than the Newtonian one in their numerical studies. Pincombe and Mazumdar~\cite{Pincombe1997} performed an analytical study for single and multiple stenosis cases, and found that the flow resistance increases with increasing stenosis height, but this increment was smaller for the Bingham blood model than the Newtonian one. Furthermore, they recommended the Bingham blood model for the analysis rather than the Casson and power-law models for small arteries. Chakravarty and Mandal~\cite{Chakravarty1997} used finite difference numerical simulations to study the blood flow through an aortic bifurcation with stenosis present in the parent artery and found that the non-Newtonian property of blood does not have any influence on the flow behaviour in the parent artery but has a significant influence on the daughter arteries. Gijsen et al.~\cite{Gijsen1999_1} performed both Laser Doppler anemometry experiments and finite element numerical simulations to investigate the influence of blood's non-Newtonian behaviour on the flow dynamics in a model carotid bifurcated artery under steady inflow conditions, wherein they found a very good agreement between the experiments and simulations. Sub-Figure~\ref{Figure5}(a) shows the schematic of their geometry, and the corresponding measurement locations are also marked in the same schematic. Sub-Figure~\ref{Figure5}(b) displays the results on axial velocity profiles (first two rows) and streamline patterns (last row), both for Newtonian and Carreau-Yasuda non-Newtonian model for blood. It can be seen that the non-Newtonian blood model exhibits a significant flattening, higher velocity gradients at the non-divider wall, and lower velocity gradients at the divider wall. In a subsequent study~\cite{Gijsen1999_2}, they conducted both experiments and simulations of unsteady flows through a large curved artery, accounting for non-Newtonian blood properties, and again observed substantial differences in the flow profiles between the Newtonian and non-Newtonian blood models. Buchanan et al.~\cite{Buchanan2000} numerically investigated the effects of three blood rheological models, namely, Newtonian, power-law, and Quemada, through a stenosed artery under various extents of pulsatile flow conditions, characterised by the Womersley number. They showed that the shear-thinning behaviour of blood has very little influence on the flow field but has a considerable effect on hemodynamic parameters evaluated at the arterial wall, such as wall shear stress and the oscillatory shear index, particularly at low Womersley numbers. Siauw et al.~\cite{Siauw2000} also compared their numerical results obtained with three rheological models of blood, namely, Casson, power-law, and Newtonian, and found no significant differences in the vortical structures of the flow field, but observed noticeable differences in the values of wall shear stress (WSS) and wall shear stress gradient (WSSG). The Casson model yielded the highest WSS values, followed by the power-law and Newtonian models, whereas a reverse trend was seen for WSSG.

\begin{figure}
    \centering
    \includegraphics[width=10cm]{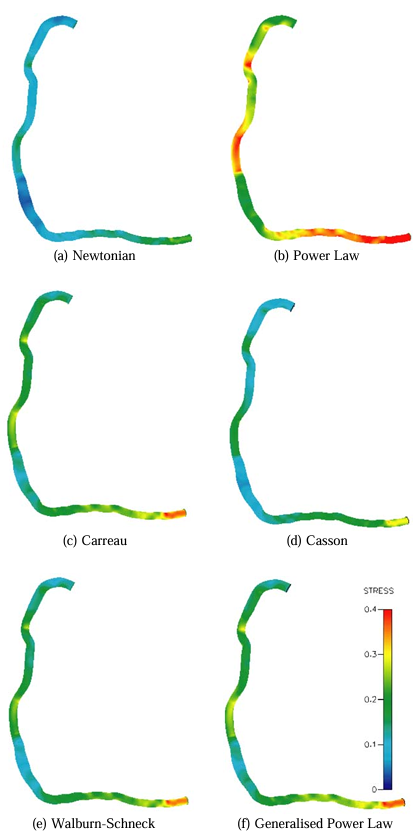}
    \caption{Distribution of wall shear stress in a coronary artery for different blood rheological models at a particular inlet velocity of 0.02 $m \,s^{-1}$~\cite{Johnston2004}.}
    \label{Figure6}
\end{figure}
Johnston et al.~\cite{Johnston2004} performed steady numerical simulations for various right coronary arteries reconstructed from bi-plane angiograms under different steady inflow conditions, considering six different rheological blood models, namely, Newtonian, power-law, Carreau, Casson, Walburn-Schneck, and generalised power law. They observed significant differences in the distribution of hemodynamic parameters such as wall shear stress, particularly at low inlet velocities, as presented in Figure~\ref{Figure6} for various blood rheological models. From this figure, the power-law model exhibits higher WSS values throughout the entire artery compared to the other models, and prominent differences are observed in the WSS distribution between the Newtonian and other non-Newtonian models. However, as the inlet velocity increased, the differences among rheological models decreased. They suggested that while the Newtonian assumption of blood rheology is good for mid-range to high shear, the generalised power-law model provides a better approximation of wall shear stress at low shear. In a subsequent study, Johnston et al.~\cite{Johnston2006} performed transient simulations and concluded that the assumption of the Newtonian model is good when predicting wall shear stress, but the non-Newtonian model is appropriate when the flow behaviour within the artery needs to be studied. Chen and Xi-Yuan~\cite{Chen2004} performed numerical simulations for a bifurcated artery incorporating the non-Newtonian property of blood using the Carreau-Yasuda model and found a significant difference in the predictions of flow behaviour and wall shear stress between Newtonian and non-Newtonian models of blood. However, they showed that when the Reynolds number is defined using a rescaled shear rate, the differences between the two models decrease. However, in a subsequent study by the same authors on pulsatile flow~\cite{Chen2006}, they found that the rescaled Newtonian blood model underestimates WSS values in low-wall-shear-rate regions and overestimates them in high-shear-rate regions. The differences were significant even after using the rescaled Newtonian blood model. They further extended their study to distal vascular graft anastomoses and found that the non-Newtonian property of blood significantly influences the flow patterns and WSS distribution within the domain~\cite{Chen2006Bipass}. For the same geometry, Vimmr and Jonášová~\cite{Vimmr2010} also found a significant influence of non-Newtonian behaviours of blood on the hemodynamics. In back-to-back studies, Abraham et al.~\cite{Abraham2005one,Abraham2005two} showed the significance of considering non-Newtonian shear-thinning blood property (which they considered using the modified Cross model) for the optimal design of arterial grafting. In particular, they observed a substantial difference in flow behaviour between the Newtonian and non-Newtonian models due to variations in the shear rate across the domain, which, in turn, affects the apparent viscosity and hence the flow behaviour. O'Callaghan et al.~\cite{OCallaghan2005} conducted numerical simulations and compared the wall shear stress distribution for a vascular bypass graft anastomosis using various rheological blood models, including Newtonian, power-law, Carreau-Yasuda, bi-exponential, cross, modified cross, and Herschel-Bulkley. They found that at high shear rates, there was almost no difference in WSS predictions among rheological models; however, at low shear rates, there was a significant difference of up to 300\%. Valencia et al.~\cite{Valencia2005} performed numerical simulations for a patient-specific model of the carotid artery with a saccular aneurysm by considering both Newtonian and non-Newtonian Herschel-Bulkley blood models. The non-Newtonian effect on wall shear stress predictions was significant in arterial regions with high velocity gradients, whereas in aneurysmal regions, predictions from Newtonian and non-Newtonian models were similar. Soulis et al.~\cite{Soulis} performed numerical simulations of the left coronary artery bifurcation and observed that the Newtonian blood model shows lower values of the wall shear stress predictions than the non-Newtonian power-law blood model. 

\begin{figure}
    \centering
    \includegraphics[width=10cm]{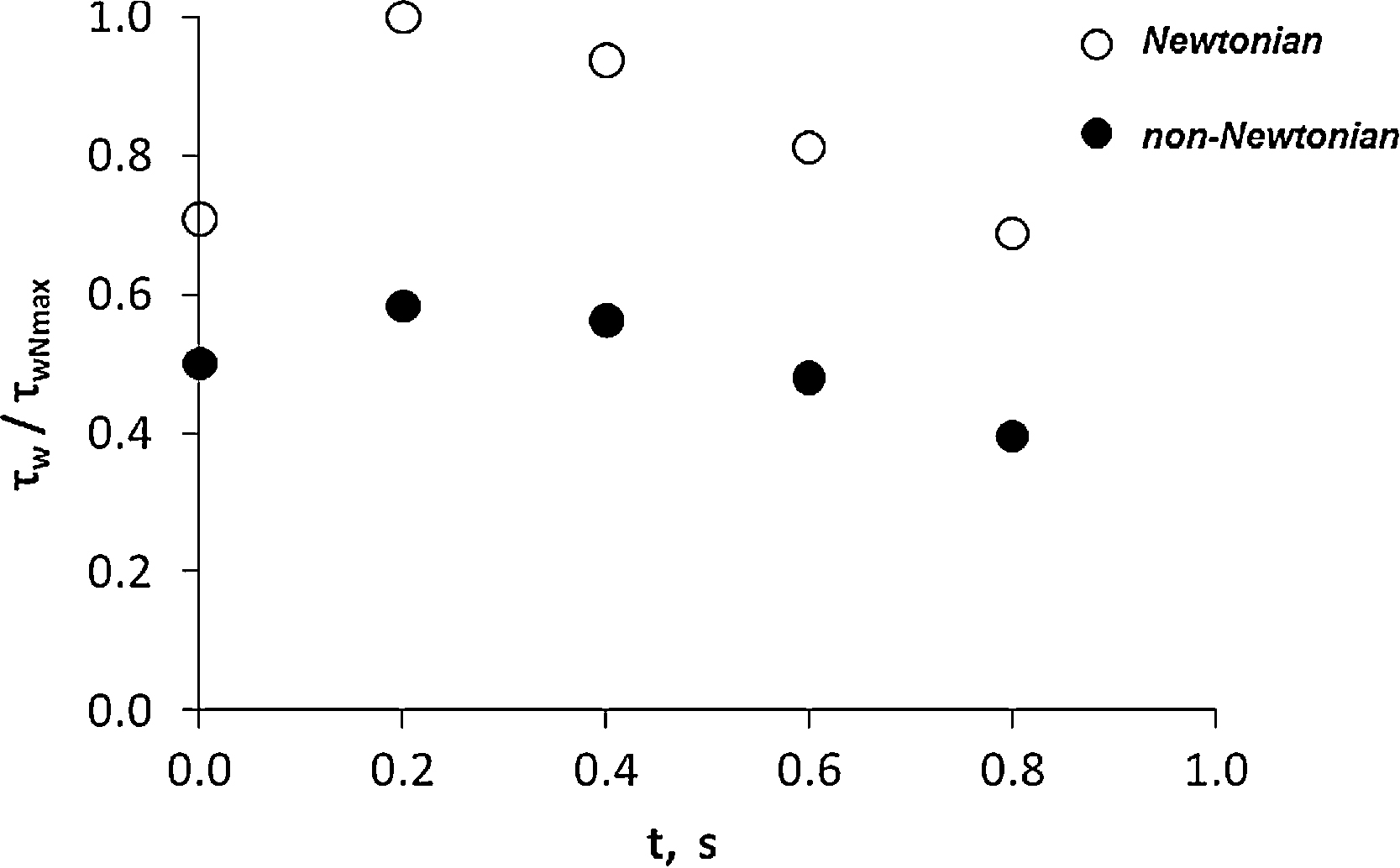}
    \caption{Experimental variation of the ratio of the wall shear stress $\tau_{w}$ obtained with a non-Newtonian blood analogue fluid to that of the maximum value obtained with a Newtonian blood analogue fluid $\tau_{WNmax}$ (normalised wall shear stress) over a pulse in a bifurcated artery~\cite{Anastasiou2011}.}
    \label{Figure6}
\end{figure}

Boyd and Buick~\cite{Boyd2007} performed two-dimensional numerical simulations for a carotid artery and showed that the Newtonian and non-Newtonian (Carreau-Yasuda) blood models exhibit small differences (less than 5\%) in the predictions of velocity and shear fields, and therefore, they concluded that in the computational hemodynamic studies, the effect of non-Newtonian behaviour can be safely neglected. Ismail et al.~\cite{Ismail2008} performed finite difference numerical simulations for a tapered overlapping stenosed artery and found that the wall shear stress values are higher for the Newtonian blood model than the non-Newtonian power-law blood model. Fan et al.~\cite{Fan2009} conducted numerical simulations for a carotid bifurcated artery with three blood rheological models, namely, Newtonian, Casson and hybrid, wherein the last model includes both Newtonian (with shear rate > 10$s^{-1}$) and Casson (for shear rate $\le 10\,s^{-1}$) depending on the range of the shear rate. They showed that the predictions for axial velocity, secondary flow, and wall shear stress obtained with the Newtonian and hybrid models were similar, whereas the Casson model produced significant differences relative to these two models. Therefore, they concluded that the Newtonian blood model is a good approximation for the hemodynamic study of a carotid artery bifurcation. Janela et al.~\cite{Janela2010One,Janela2010two} performed numerical simulations of a stenosed artery, considering fluid-structure interaction between the artery wall and blood, and found a noticeable difference in the displacement of the artery wall between Newtonian and non-Newtonian rheological assumptions of blood. Park et al.~\cite{PARK2010} performed both numerical simulations and in-vitro micro particle image velocimetry experiments to study the hemodynamics in a stenosed artery and found that the rheological behaviour of blood presented by several models hardly influences the velocity field near the artery wall, where the shear rate is high; however, at the centre where the shear rate is low, it has a strong influence. Razavi et al.~\cite{Razavi2011} conducted numerical simulations of a symmetric stenosed carotid artery using different rheological models of blood and observed that the power-law blood model deviates significantly from the Newtonian blood model in its predictions of wall shear stress and the velocity field. For a 3D bifurcating abdominal aortic aneurysm model, Ma and Turan~\cite{Ma2011} conducted numerical simulations for physiologically realistic pulsatile flow and found that the Newtonian blood model under-predicts the maximum values of WSS compared to the non-Newtonian Carreau-Yasuda model. Morbiducci et al.~\cite{Morbiducci2011} conducted numerical simulations for a carotid bifurcated artery and concluded that the Newtonian assumption of blood rheology could be reasonable for predicting bulk flow metrics with differences less than 10\% obtained with the non-Newtonian rheological model. An experimental study was conducted by Anastasiou et al.~\cite{Anastasiou2011} to study the hemodynamics of a blood analogue fluid through a small bifurcated artery. Their results revealed that the Newtonian blood analogue overestimates WSS values more than the non-Newtonian one, and hence they concluded that one cannot neglect the importance of blood's non-Newtonian rheology in predicting hemodynamic parameters, particularly for flow through small arteries. Their subsequent numerical study also further established this finding~\cite{Kanaris2012}. In contrast, Xiang et al.~\cite{Xiang2011} showed in their numerical simulations carried out for the carotid artery aneurysm that the Newtonian blood model could underestimate WSS values compared to non-Newtonian blood models, particularly in the complex-shaped saccular aneurysm, where low shear rate prevails, which facilitates the thrombus formation and rupture of the aneurysm. Therefore, the assumption of a Newtonian blood model may underestimate the occurrence of these events.  Cherry and Eaton~\cite{Cherry2013} conducted numerical simulations for straight and curved arteries and found significant differences in the hemodynamics when considering blood as simple Newtonian and shear-thinning non-Newtonian fluids. For instance, in both straight and curved arteries, the velocity profile was flattened, and the shear-thinning blood model exhibited a higher pressure drop than the Newtonian one. Furthermore, in the curved artery, the secondary flow strength was diminished when the blood was assumed to be shear-thinning. For a 3-dimensional idealized femoral artery tree, Weddell et al.~\cite{Weddell2015} showed that the non-Newtonian Carreau-Yasuda model predicted higher WSS values and pressure drop than the Newtonian one. Doost et al.~\cite{Doost2016} conducted numerical simulations of a patient-specific left ventricle (LV) using different non-Newtonian rheological blood models, as well as the Newtonian model. They found that the magnitude and number of small vortices, as well as the maximum WSS values, varied across rheological models. Carty et al.~\cite{Carty2016} predicted the hemodynamics through an intracranial aneurysm using both Newtonian and non-Newtonian rheological blood models, and suggested that there should be a cap in the low and high shear rate viscosity values available in the non-Newtonian models. They found similar maximum WSS values for both Newtonian and non-Newtonian models; however, the regions of low WSS differed between the two models.  Najjaari and Plesniak~\cite{Najjari2016} conducted experiments to analyse the vortical structures in a curved artery using both Newtonian and non-Newtonian blood analogue fluids, and found hardly any difference in the secondary flow structure between the two fluids. They suggested that it might be due to the large artery size, which is not allowing the shear-thinning properties to be prominent. Soares et al.~\cite{Soares2017} performed numerical simulations to study the hemodynamics of an abdominal aorta bifurcation using both Newtonian and non-Newtonian Walburn-Schneck’s models. Throughout the cardiac cycle, they found higher wall shear stress values with Walburn-Schneck’s model than with the Newtonian one. Iasiello et al.~\cite{Iasiello2017} concluded that the Newtonian model is valid without any particular loss in terms of accuracy, particularly at higher velocities when comparing with other non-Newtonian rheological models for their simulations on an aorta-iliac bifurcation. 

\begin{figure}
    \centering
    \includegraphics[width=10cm]{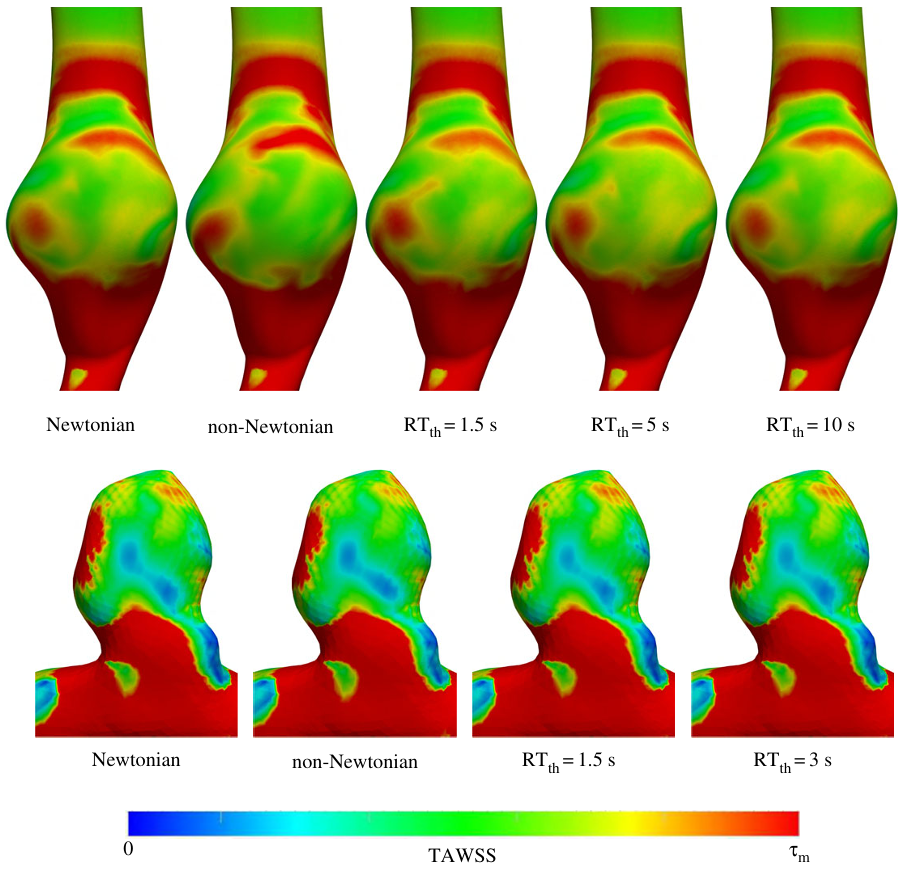}
    \caption{Distribution of time-averaged wall shear stress (TAWSS) for different blood models for the abdominal aortic aneurysm (top row) and cerebral aneurysms (bottom row)~\cite{Arzani2018}. Here, the results presented in the second column were obtained with the Carreau-Yasuda non-Newtonian model, whereas the results obtained with different residence times are presented by the label $RT_{th}$.}
    \label{Figure7}
\end{figure}

Arzani~\cite{Arzani2018} proposed a novel rheological model in which the non-Newtonian shear-thinning behaviour of blood was assumed to be not only a function of shear rate but also of residence time (RT) in the low-shear-rate regime. This concept was based on the experimental fact that the shear-thinning behaviour of blood is a function of red blood cell aggregation and rouleaux formation, which require sufficient residence time in a low-shear-rate regime. With this newly proposed rheological model, the author conducted numerical simulations of blood flow through cerebral and abdominal aortic aneurysms, along with Newtonian and non-Newtonian Carreau-Yasuda models for blood. Figure~\ref{Figure7} shows the surface distribution of time-averaged wall shear stress (TAWSS) obtained with different blood rheological models for the two geometries. One can see that the WSS distribution looks almost indistinguishable for different rheological models of blood. Therefore, the author suggested that the traditional shear-thinning non-Newtonian models, such as the Carreau-Yasuda model, used in the literature exaggerate the non-Newtonian behaviour of blood. Furthermore, the author's proposed RT-based non-Newtonian model showed a significant reduction in shear-thinning effects and exhibited hemodynamic behaviours qualitatively and quantitatively similar to those of the simple Newtonian model.            

\begin{figure}
    \centering
    \includegraphics[width=17cm]{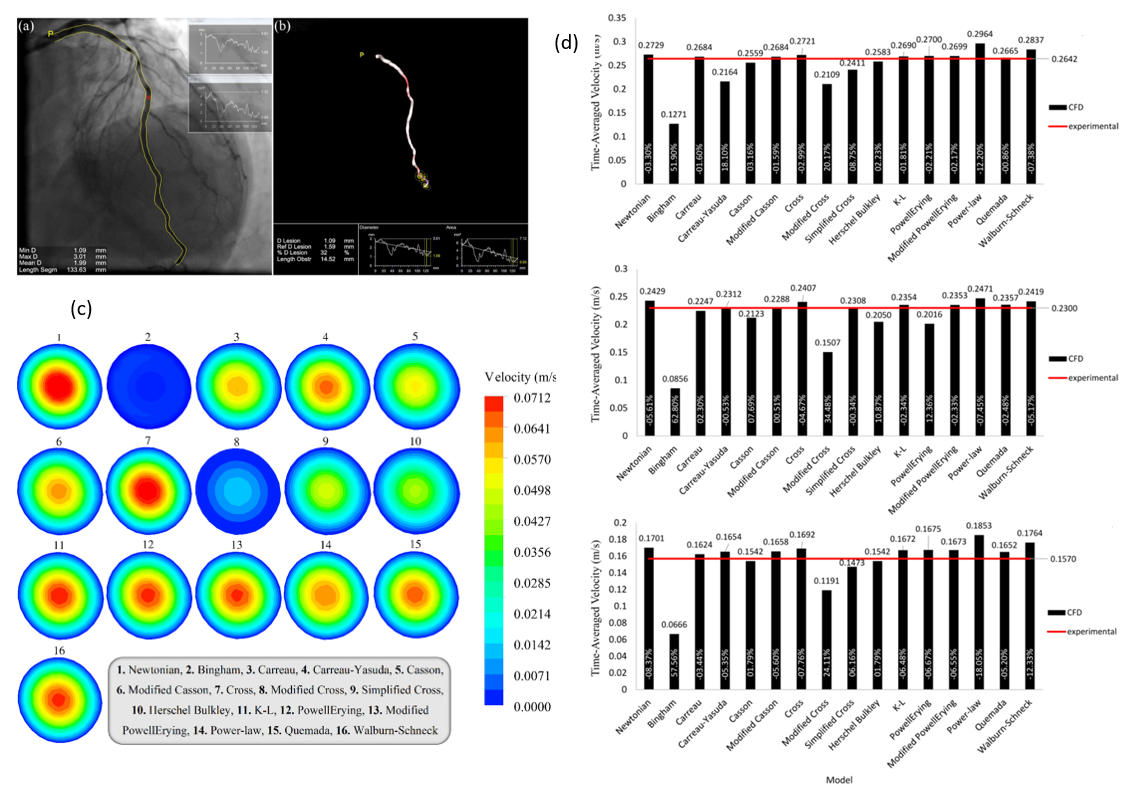}
    \caption{(a) Reconstruction of the three-dimensional model of the coronary artery with stenosis from the coronary angiography with geometrical data and (b) the corresponding three-dimensional reconstructed model. (c) Comparison of the velocity field distribution in the middle of the artery with mild stenosis obtained with different rheological models of blood. (d) Comparison of the time-averaged velocity at a point in the middle of the artery between clinical measurements and CFD simulations obtained with different rheological models of blood for different stenosis conditions, namely, mild, moderate and severe~\cite{Abbasian2020}.}
    \label{Figure8}
\end{figure}
 
Abbasian et al.~\cite{Abbasian2020} were probably the first to quantitatively compare the numerical results obtained with various rheological models of blood and in vivo experiments. Their CFD simulations were performed using three-dimensional patient-specific models of the diseased left anterior descending (LAD) coronary arteries with varying degrees of stenosis severity, which were reconstructed from 3D quantitative coronary angiography, as schematically shown in Sub-Figure~\ref{Figure8}(a). Sub-Figure~\ref{Figure8}(b) demonstrates the velocity distribution at a mid plane obtained with Newtonian and different non-Newtonian models of blood, and it can be seen that there is a significant difference present obtained with different models under the flow conditions. A quantitative comparison of the time-averaged velocity value at a mid-point within the artery between the in vivo clinical measurements and CFD simulations with different rheological models of blood is presented in sub-Figure~\ref{Figure8}(c) for different stenosis severity, namely, mild, moderate, and severe. This comparison suggests that the non-Newtonian models, particularly the Carreau, modified Casson, or Quemada models, match the corresponding clinical measurements exceptionally well compared to Newtonian and other non-Newtonian models. Lopes et al.~\cite{Lopes2020} performed numerical simulations for a carotid artery with fluid-structure interaction and different rheological models of blood, and they found that the WSS values obtained with the Carreau model are 13\% higher than the Newtonian model. Abugattas et al.~\cite{Abugattas2020} conducted three-dimensional numerical simulations for a carotid artery geometry and predicted the wall shear stress using three rheological models of blood, namely, power-law, Cross, and Carreau-Yasuda. They found that the prediction using the power-law model was lower than that of the Cross and Carreau-Yasuda models. Liu et al.~\cite{Liu2021} performed simulations for the intracranial atherosclerotic stenosis (ICAS) artery using Newtonian and two non-Newtonian blood models (Casson and Carreau-Yasuda) and found that the difference in the prediction of Wss is insignificant in the high shear-rate regions, but it is significant in the low-shear rate regions. Thondapu et al.~\cite{Thondapu2022} predicted significantly higher values of the patient-specific coronary endothelial shear stress (ESS) calculations using the non-Newtonian Quemada model than those using the Newtonian model. Samaee et al.~\cite{Samaee2022} conducted both experiments and numerical simulations to investigate the effect of Newtonian and non-Newtonian blood rheology (Carreau model) on hemodynamic assessment in a carotid bifurcation. They fed the flow and pressure pulses from their experimental model into their numerical model, which included fluid-structure interaction and evaluated various clinically important parameters, such as the oscillatory shear index, vorticity, and Von-Mises stress. Once again, they found a significant difference in the predictions of these parameters by the Newtonian and non-Newtonian models. De Nisco et al.~\cite{DeNisco2023} carried out simulations of 144 right coronary arteries with different stenosis degrees, which were reconstructed from angiography by considering Newtonian and non-Newtonian Carreau blood models. They concluded that blood rheology has a negligible effect on hemodynamic parameters such as WSS and helical flow profiles.    
 
So far, we have reviewed the influence of blood's shear-thinning rheological behaviour on the hemodynamics in various arteries. However, as discussed earlier, blood also exhibits viscoelastic rheological behaviour. The influence of this particular rheological behaviour, along with shear-thinning behaviour, has not been investigated to the same extent as that of blood's generalised non-Newtonian fluid behaviour. Among the very few studies, Sharp et al.~\cite{SHARP1996} conducted an analytical study to investigate the effect of blood viscoelasticity in a straight rigid artery using the Maxwell constitutive model compared with the Newtonian blood model. They found almost negligible differences in the WSS values and velocity field between the two models. Leuprecht and Karl~\cite{LEUPRECHT2001} performed numerical simulations of large arteries comprising a stenosed artery and a 90-degree curved artery under steady-flow conditions. They used a modified Oldroyd-B viscoelastic model that accounted for both the viscoelasticity and shear-thinning rheological behaviours of blood. Their simulations revealed that blood viscoelasticity has little effect on flow dynamics in a 90-degree curved artery, whereas it has a noticeable effect on flow through a stenosed artery. Therefore, they concluded that the influence of blood viscoelasticity on the flow dynamics depends on the artery's geometric configuration. In contrast, Rojas~\cite{Rojas2006} found that the blood viscoelasticity modelled by the six elements of the Maxwell model (multimode Maxwell model) has a significant effect on the values of the oscillatory shear index in flows even through a straight artery. In particular, the viscoelastic blood model predicted OSI values that were almost 12\% higher than those of the Newtonian model. Similarly, Elhanafy et al.~\cite{Elhanafy2019} also found higher values of the wall shear stress and pressure on the wall when blood viscoelasticity and shear-thinning were both considered compared to the simple Newtonian assumption for the flow through an abdominal aortic aneurysm under steady flow conditions. They modelled blood viscoelasticity using the Oldroyd-B model, whereas the shear-thinning was modelled using the Carreau-Yasuda model. Using the same approach in the constitutive modelling of blood, Bilgi and Kunt~\cite{Bilgi2020} performed numerical simulations for the hemodynamics through an artery with an axisymmetric aneurysm present in it under pulsatile flow conditions and fluid-structure interaction between the artery wall and blood. They also observed higher wall shear stress and von Mises stress on the arterial wall when blood viscoelasticity was accounted for. Furthermore, the vortical structure formed within the aneurysm was found to be strongly influenced by viscoelasticity and shear-thinning compared with the simple Newtonian assumption. For instance, they showed that blood viscoelasticity and shear-thinning lead to the formation of secondary vortices within the aneurysm, alongside the primary vortex, whereas these secondary vortices were absent when a simple Newtonian blood model was considered. Pinto et al.~\cite{Pinto2020} performed three-dimensional numerical simulations for patient-specific right coronary arteries with and without stenosis present in them using three different rheological models for blood, which account for blood viscoelasticity, namely, the generalised Oldroyd-B model, the multimodal Giesekus model, and the multimodal simplified Phan-Thien-Tanner (sPTT) model. Their simulations revealed that the wall shear stress values decreased (almost by half) when considering both shear-thinning and viscoelasticity of blood compared to only shear-thinning (modelled using the Carreau model). 

\begin{figure}
    \centering
    \includegraphics[width=12cm]{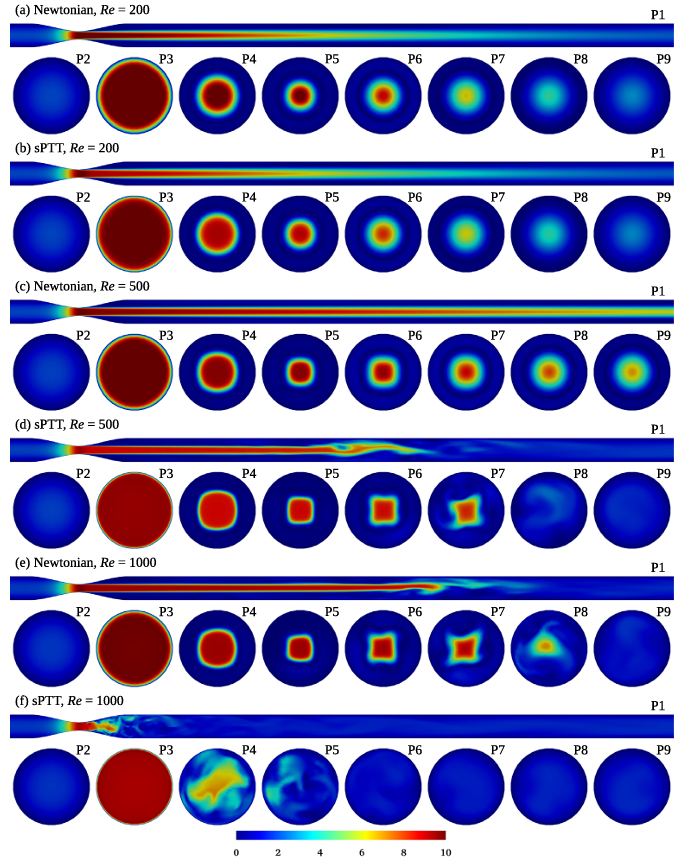}
    \caption{Distribution of velocity field inside a stenosed artery at different flow strengths (quantified in terms of the Reynolds number), both for Newtonian and sPTT rheological models of blood under steady flow conditions~\cite{Chauhan2026}. The visualisation is present at different planes normal to the flow direction (planes P2 to P8) and parallel to the flow direction (plane P1).}
    \label{Figure9}
\end{figure}

Bodnar et al.~\cite{Bodnr2011} conducted numerical simulations for the flow through a stenosed artery and a carotid bifurcated artery under steady flow conditions, considering four scenarios of blood rheological behaviour: simple Newtonian, purely viscoelastic (Oldroyd-B model), purely shear-thinning (Cross model), and combined viscoelastic and shear-thinning (generalised Oldroyd-B model). They concluded that the shear-thinning effect of blood is more prominent in modulating flow variables, such as velocity and pressure, and the wall shear stress distribution, compared with the viscoelastic effect. Recently, Chauhan and Sasmal~\cite{Chauhan2021} carried out two-dimensional axisymmetric numerical simulations for the flow through a stenosed artery using the multimode sPTT constitutive model for blood under both steady and pulsatile flow conditions. They used realistic values for the model parameters by fitting the experimental rheological data for real and whole blood from steady-shear and small-amplitude oscillatory-shear experiments. Significant differences both in the velocity field and the WSS distribution between the sPTT and the Newtonian model of blood were observed in their study. For instance, they observed several small-scale vortical structures within the stenosed geometry under both steady and pulsatile flow conditions, resulting in a more chaotic flow in the sPTT model than in the Newtonian one. Furthermore, the predicted pressure drop across the stenosis and WSS values were higher for the Newtonian model. Subsequently, they extended their study to three-dimensional stenosed geometry and observed phenomena similar to those seen in the two-dimensional case~\cite{Chauhan2026}. Figure~\ref{Figure9} shows a representative plot of velocity field distribution at different planes inside the artery at different Reynolds numbers under steady flow conditions, both for Newtonian and sPTT rheological models of blood. From this figure, it can be seen that a central high-velocity-magnitude jet zone forms in the stenosed area, which is shorter, more distorted, and broken in the sPTT model compared to the Newtonian one. This suggests that the flow field becomes more chaotic within the artery when blood is treated as viscoelastic and shear-thinning rather than simple Newtonian. Considering blood thixotropy, viscoelasticity and viscoplasticity, a pulsatile blood flow through three-dimensional rigid aneurysmal geometries was investigated by Giannokostas and Dimakopoulos~\cite{Giannokostas2023} using the thixo-elasto-viscoplastic (TEVP) constitutive model, with model parameters calibrated for blood from healthy subjects. A systematic parametric analysis was performed by considering sinusoidal pulsatile waveforms with varying frequencies and amplitudes. They observed that increasing the pulse frequency reduces the temporal variation of the structural parameter and promotes the formation of a recirculation zone within the aneurysmal dome. A substantially higher wall shear stress was developed near the aneurysm mouth, which may contribute to progressive enlargement of the aneurysmal sac. Interestingly, the WSS exhibited pronounced spatial oscillations and higher amplitudes along the parent vessel section adjacent to the aneurysm, where flow instability developed. Increasing the pulse amplitude made the blood structurally weaker, thereby facilitating flow. Nevertheless, the WSS within the aneurysm remained relatively low, a condition that may favour aneurysm growth and potentially rupture, as low wall shear stresses have been associated with these pathological processes. However, they did not provide any comparison with either Newtonian or generalised non-Newtonian fluid models, so it was not possible to determine how blood's complex rheological behaviours influence hemodynamics.

\begin{figure}
    \centering
    \includegraphics[width=12cm]{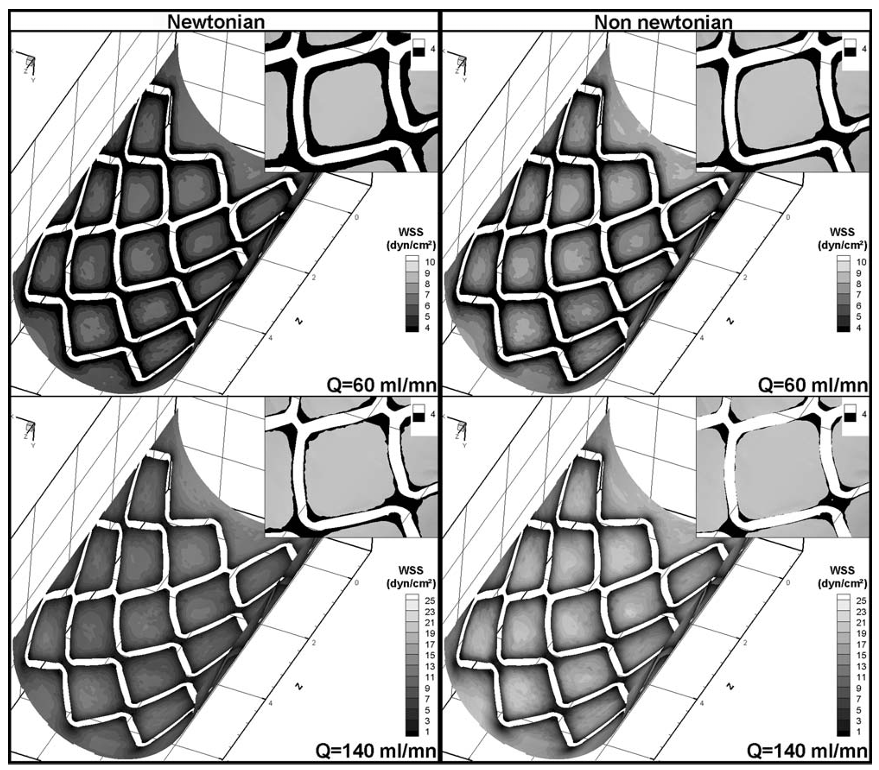}
    \caption{Distribution of wall shear stress on the wall of a coronary stent artery obtained with Newtonian and non-Newtonian blood models at two different flow rates inside the artery~\cite{Benard2006}. Here, the dark black colour indicates low wall shear stress regions, whereas the grey colour indicates high wall shear stress regions.}
    \label{Figure10}
\end{figure}

\begin{figure}
    \centering
    \includegraphics[width=12cm]{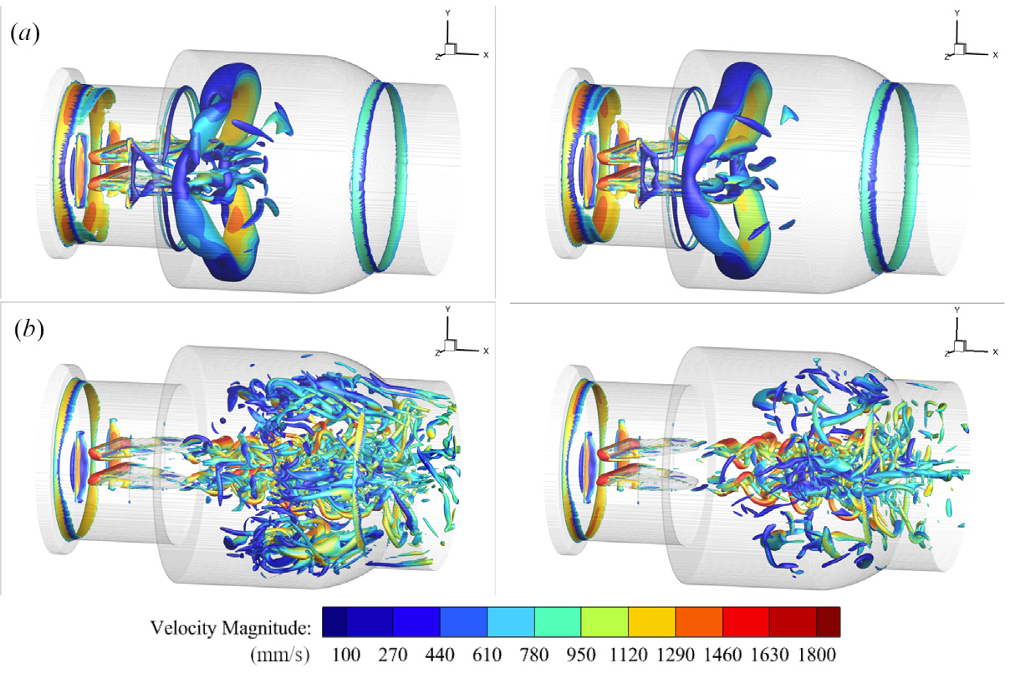}
    \caption{Q-criterion for vortex structure distribution. Here, the first and second columns represent the results for the Newtonian and non-Newtonian models, respectively. (a) mid acceleration, t = 0.1 s; (b) just before peak acceleration, t = 0.18 s, of a cardiac cycle~\cite{Sarkar2024}.}
    \label{Figure11}
\end{figure}

\section{\label{section5}Impact of blood rheology on cardiovascular device flows}
In this section, we present a brief discussion on the importance of blood rheology in the hemodynamics of various cardiovascular devices. We focus in particular on two devices, namely, prosthetic mechanical heart valves and stents. Seo et al.~\cite{Seo2005} conducted a numerical study for the blood flow through a straight and curved artery with a stent present in it under both steady and pulsatile flow conditions. They observed flow disturbances around and downstream of the stent, which increased with Reynolds number and stent height. Their study showed that the non-Newtonian shear-thinning property of blood, modelled by the Carreau model, has little effect on the wall shear stress values obtained with the Newtonian model. However, they observed an 8\% reduction in the flow recirculation zone when the blood's non-Newtonian rheology was accounted for. Simulation studies by Ohta et al.~\cite{Ohta2005} for an intracranial aneurysm with a stent revealed that the wall shear stress was reduced and the local viscosity was increased inside the aneurysm because of the presence of the stent. The increase in local viscosity within the aneurysm was attributable to their use of the non-Newtonian Cross model for blood. For a similar geometry, Mezali et al.~\cite{Mezali2022} also found that the non-Newtonian properties of blood provide significantly higher values of the wall shear stress within the aneurysm compared to those obtained with the simple Newtonian model. McCarthy et al.~\cite{McCarthy2025} showed that the difference in the time-averaged wall shear stress values between the Newtonian and non-Newtonian (Carreau-Yasuda) models is greatest near the stent in their simulations for a coronary stent artery. A similar finding was reported by Benard et al.~\cite{Benard2006}. Figure~\ref{Figure10} shows the distribution of wall shear stress for the two rheological models of blood at two different flow rates obtained in their simulations. It can be seen that the regions of low wall shear stress present near the strut are smaller when the non-Newtonian shear-thinning property of blood is present. As the flow rate increases, this effect becomes even more pronounced. Ahadi et al.~\cite{Ahadi2023} carried out numerical simulations using three rheological models, namely Newtonian, non-Newtonian power-law, and Carreau, and found that the Carreau model yields higher wall shear stress values than the other two models in a coronary stent artery. Kasaeinia et al.~\cite{Kasaeinia2025} investigated the effects of the strut shape of the stent and the blood rheological model on the hemodynamics in a coronary stent artery. They observed that for a square-shaped strut, the flow fields and hemodynamic parameters, such as wall shear stress and pressure drop, differed noticeably between the Newtonian and non-Newtonian blood models. In contrast, for a semi-circular-shaped strut, the difference was negligible. Boniforti et al.~\cite{Boniforti2024} conducted numerical simulations for an intracranial aneurysm with a flow diverter stent (FDS) using Newtonian and non-Newtonian (Carreau and Carreau-Yasuda) rheological blood models and found that the Newtonian model overestimated the relative residence time (RRT) and so the consequent aneurysm healing compared to the non-Newtonian model. Lei et al.~\cite{Lei2015} studied the effect of blood rheology on flow dynamics and mass transfer in a thoracic aortic aneurysm before and after a virtual stent operation. They found that the Newtonian assumption of blood rheology underestimated the wss values compared to the non-Newtonian Carreau model. Furthermore, they observed that the non-Newtonian property of blood significantly influences LDL and oxygen transport within the aneurysm. Walker et al.~\cite{walker2012quantification} performed an in vitro experiment to investigate the influence of Newtonian and non-Newtonian analogue fluids on the quantification of hemodynamics parameters downstream of a stent. They found that the spatial coverage of flow reversal and the oscillatory shear index downstream of the stent were reduced appreciably in non-Newtonian blood analogue fluid compared to Newtonian blood analogue fluid. Wang et al.~\cite{Wang2012} conducted a simulation study to analyse the distribution of drug concentration from a drug-eluting stent using Newtonian and non-Newtonian Carreau blood models. They found no significant differences in the distribution of drug concentration across most vessel walls and in area-averaged surface drug concentration between the two models. Huang et al.~\cite{Huang2013} performed simulations for a stented cerebral aneurysm using Newtonian and non-Newtonian Casson blood models. For an aneurysm with a larger neck size and high stent porosity, they found negligible differences between the two models. However, for a smaller neck size and low stent porosity, the differences were significant, with the Casson model underestimating the WSS values.

Few studies have also been carried out in the literature to investigate the influence of blood rheological behaviours on the hemodynamics past a prosthetic mechanical heart valve. For instance, De Vita et al.~\cite{DeVita2015} conducted three-dimensional numerical simulations of a bileaflet mechanical aortic heart valve, considering blood as both a Newtonian and a Carreau-Yasuda shear-thinning non-Newtonian fluid. They found that the blood rheological behaviour has little influence on valve dynamics and the transvalvular pressure drop. In contrast, the hemolysis index (HI), signifying the blood damage caused by the abnormal stresses, which in turn is induced by the mechanical heart valve, was found to be 20\% larger for the shear-thinning blood model than the simple Newtonian model. Using the same Carreau-Yasuda shear-thinning model of blood, Sarkar et al.~\cite{Sarkar2024} also recently conducted a numerical study and again found blood damage to be 21\% higher than that of the Newtonian model. Not only did they find a substantial difference in the blood damage index, but they also found substantial differences in the flow field, including vortex structures. Figure~\ref{Figure11} depicts the Q-criterion for vortex structure distribution colored by the corresponding velocity magnitude for both Newtonian and non-Newtonian models for two distinct time instances of a cardiac cycle. They found that the non-Newtonian blood model exhibits greater stability of the vortex ring formed in the sinus region than the Newtonian blood model. Furthermore, the non-Newtonian blood model exhibited fewer small-scale vortex structures than the Newtonian model. Overall, they concluded that the non-Newtonian blood model provides a more organised flow field than the Newtonian blood model. Chen et al.~\cite{Chen2022} conducted the numerical analysis of the hemodynamics of a mechanical heart valve by considering several non-Newtonian blood models along with the Newtonian model, and found that the Casson model provides around 70.34\% higher wall shear stress values compared to the Newtonian model. They further pointed out that the narrowing of the hinge region during heart valve leaflet movement results in a low flow rate, wherein blood rheology significantly influences the hemodynamics. Pohl et al.~\cite{Pohl1996} performed an in vitro experiment to investigate the influence of blood rheological behaviours on the flow dynamics of a Björk-Shiley heart valve. They considered several Newtonian (water, glycerol solutions) and non-Newtonian (polyacrylamide, xanthan gum) blood analogue fluids for the experiments. Their results revealed that the closing time and closing volume are not influenced by the blood rheological behaviour. However, the leakage flow and leakage volume were greatly influenced by the rheological behaviour of blood. Based on their study, they concluded that the non-Newtonian rheological behaviour of blood should be considered when modelling and testing prosthetic valve performance, not only by accounting for the shear-rate-dependent viscosity of blood but also for viscoelasticity and thixotropy. Very recently, Lupi et al.~\cite{Lupi2026} performed large-scale numerical simulations of the left heart with aortic valve and found that the non-Newtonian rheological behaviours of blood hardly affect the blood pressure; however, substantial differences were seen for the wall shear stress and hemolysis index between the Newtonian and non-Newtonian models of blood. In back-to-back studies, Chauhan and Sasmal~\cite{Chauhan2024,Chauhan2024Dis} carried out a detailed investigation into the hemodynamics of a bileaflet mechanical heart valve under both fully functional and various dysfunctional conditions, considering both Newtonian and non-Newtonian blood models, such as the power-law and Casson models, and also under both steady and pulsatile flow conditions. They observed that at low Reynolds numbers, differences in hemodynamic parameters such as wall shear stress and pressure drop were greater and gradually diminished as the Reynolds number increased under steady flow conditions. For pulsatile flow conditions, the Casson model predicted higher wall shear stress and blood damage values than the power-law and Newtonian models. Furthermore, they showed that for a dysfunctional mechanical heart valve, the difference in the prediction of clinically important parameters increased between the Newtonian and non-Newtonian blood models~\cite{Chauhan2024Dis}.

\section{\label{section6}Discussion and future directions}
At the outset of the discussion, it should be firmly accepted that blood is not a simple Newtonian fluid but rather a complex non-Newtonian fluid due to its multiphase nature, arising from the presence of a liquid-phase plasma (almost 55\%) and suspended cellular components (almost 45\%). As already discussed in section~\ref{section2}, blood exhibits several non-linear rheological behaviours ranging from shear-thinning to thixotropy. After a comprehensive review of the hemodynamics of arterial flows and certain cardiovascular devices, the major outcomes should be thoroughly discussed before concluding on how different blood rheological behaviours influence the overall hemodynamics. This will also set the platform for future research directions.  
\begin{itemize}
    \item First of all, for arterial hemodynamics, a contradictory outcome is presented on the predictions of several clinically important parameters, such as wall shear stress and pressure drop, when Newtonian and non-Newtonian blood models have been used. Some studies predicted that the Newtonian blood model yields higher WSS and pressure drop values than the non-Newtonian model~\cite{shukla1980effects,chakravarty1987effects,nakamura1988numerical,Wells1996}, whereas others reported the opposite trend~\cite{chaturani1985study,cho1991effects,luo1992non,Das1995}. Furthermore, some studies have shown that there is hardly any difference present in predictions for clinically relevant parameters~\cite{baaijens1993numerical,Arzani2018}. Therefore, the question is which prediction is correct and which is wrong. The answer lies in how different constitutive models for blood have been utilised to fit blood's experimental rheological behaviour, and under which circumstances those models have been used. For instance, when using the Newtonian blood model, one must use a fixed value of the blood viscosity in the analysis. Most studies used a value of 3.5 mPa.s for the infinite-shear viscosity $(\eta_{\infty})$ when blood is undergoing steady shear flow. This is the value of apparent viscosity blood would exhibit when the shear rate is above 100 $s^{-1}$. However, spatial and temporal variations in shear rate will occur within an artery, depending on its size and the cardiac cycle phase. The size of the aorta is in the range of 20-20 mm, whereas it is in the range of 2-5 mm for coronary arteries~\cite{Muneeb2023}, and as a result, the distribution of shear rate in these arteries will be different. Furthermore, during the peak systolic stage of the cardiac cycle, the velocity is at its maximum, whereas it is at its minimum during the diastolic stage, which in turn leads to temporal variation in the shear rate and, hence, the apparent viscosity of blood. For instance, the range of average shear rate in the coronary artery can be between 100-300 $s^{-1}$, whereas that in the aorta can be between 40-150 $s^{-1}$~\cite{Panteleev2021,Doriot2000}. On top of these, for a stenosed artery or a bifurcated artery, a spatial variation in the shear rate will again be there. For instance, in a stenosed artery, the shear rate will be high around the stenosis due to the high velocity magnitude in this region, because of the narrowing of the flow area and obeying the mass conservation principle. In contrast, in the recirculation region downstream of the stenosis, the shear rate will be low. Similarly, in a bifurcated artery, there will be a spatial distribution of shear rate due to the formation of recirculation zones upstream or downstream of the bifurcation. The clinically important parameter, such as wall shear stress, at a point is calculated as $\tau_w = \eta\dot{\gamma}_w$ where $\dot{\gamma}_w$ is the wall shear rate at that point and $\eta$ is the viscosity at the same point. For a Newtonian blood model, $\eta$ remains constant at 3.5 mPa.s, but for a shear-thinning non-Newtonian blood model, the viscosity (better to say apparent viscosity) itself would be a function of the local shear rate. In particular, the apparent viscosity decreases, leading to lower WSS values. However, this reduction in apparent viscosity increases the velocity gradient, leading to larger wall shear rates and, in turn, higher wall shear stress. This suggests that there is a competitive influence between these two factors, which ultimately determines the WSS values for a shear-thinning constitutive model of blood. For a situation when the shear rate exceeds 100 $s^{-1}$ or so everywhere in the flow domain, there should be no difference in the prediction between the Newtonian and non-Newtonian blood models. However, if it falls below that value at any point in the cardiac cycle or in the domain, one can expect differences between the two models. It is therefore always recommended to use the non-Newtonian shear-thinning model to capture the spatial and temporal variations of wall shear stress accurately, rather than the simple Newtonian blood model. Of course, there will be additional computational cost, as one has to calculate the apparent viscosity at each cell of the computational domain to update the viscous stress in the momentum equation. Furthermore, one may require more cells for the shear-thinning model than for the constant-viscosity Newtonian model to capture the velocity gradient, since the shear-thinning model exhibits a steeper velocity gradient. If this extra computational cost is not an issue, one should definitely utilise the non-Newtonian shear-thinning constitutive model for hemodynamic modelling rather than the simple Newtonian model. 
    
    \item Several inelastic generalised Newtonian fluid models have been utilised for flow modelling of blood, both for arterial and cardiovascular device flows, and some of them are briefly mentioned in section~\ref{section3}. These models have their own advantages and disadvantages. For instance, the power-law is the simplest formula that is easy to handle and has only two fitting parameters. As a result, this model could be easily fitted to experimental rheological data for blood. However, the limitation of this model is that it can not predict the limiting values of the apparent viscosity exhibited by a fluid at very high (infinite-shear rate viscosity, $\eta_{\infty}$) or low shear (zero-shear rate viscosity, $\eta_{0}$) rates. From its equation~\ref{eq:powerlaw}, it can be seen that the apparent viscosity will continue to decrease with the increasing shear rate values, whereas it will continue to increase with the decreasing shear rate values for a shear-thinning fluid. Therefore, according to this model, in the limits of $\dot{\gamma} \rightarrow 0$ and $\infty$, the zero-shear and infinite-shear rate viscosities for a shear-thinning fluid like blood tend to be infinite and zero, respectively, which is unrealistic. Although Figure... shows that the power-law model fits the data over the entire range of shear rates considered, a situation can arise in the computational domain in which the shear rate becomes very high or very low, leading to unrealistic apparent viscosity values in the modelling. For instance, in cardiovascular device flows, nearly stagnant regions may form in the vicinity of stent struts or the hinge of a mechanical heart valve~\cite{Benard2006,yun2012numerical,ellis1996velocity}. It is better to avoid such a situation, as it may lead to divergence in the simulations and to unrealistic, inaccurate results. Therefore, the Carreau-Yasuda or Cross model can be used more effectively, as it includes the limiting values of viscosity at high and low shear rates in its expression. 
    
    \item The role of blood viscoelasticity on the hemodynamics has been investigated to a very limited extent, as compared to that performed for the shear-thinning behaviour of blood. Although some studies are available on arterial flows, almost no investigations are available on cardiovascular devices dealing with blood flow, such as stents or mechanical heart valves. The bigger question is whether blood viscoelasticity should be considered relative to shear-thinning behaviour in the analysis. Figure~\ref{Figure3} demonstrates that blood indeed possesses viscoelasticity to a certain extent. However, the value of $G'$ (which quantifies the elasticity) remains always lower than the value of $G"$ (which signifies the viscous property of blood) over a large range of frequency. This naturally suggests that the viscous property, or more appropriately, viscous dissipation, is more significant than the storage of elastic energy in blood. Blood viscoelasticity may still be relevant in processes that can be quantified by a nondimensional number called the Deborah number (De)~\cite{reiner1964deborah}. It is defined as the ratio of two time scales, namely, the blood relaxation time $(\lambda)$ and the flow time scale $(T)$, i.e., $De = \frac{\lambda}{{T}}$. When this number is much smaller than 1, the viscous response of blood will dominate. However, a value of order one and higher will exhibit strong elastic effects of blood on the hemodynamics. For a pulsatile flow occurring during the cardiac cycle, the flow time scale can be replaced with the frequency $\omega$, so the Deborah number becomes $De = \lambda \omega$. For cardiac flow, the Deborah number typically ranges from 0 to 20~\cite{beris2021recent}. Therefore, one can see that at certain instances, the elastic effects of blood can become very significant in the analysis, and they cannot be ignored. Blood elasticity becomes increasingly important in arterial and cardiovascular device flows as pulsation frequency or relaxation time increases. In cases such as intense physical exercise, cardiac frequency increases, and the frequency-based Deborah number likewise increases; consequently, the influence of blood elasticity on flow may become substantial, as the blood microstructure, such as RBCs, has less time to relax after undergoing deformation. Furthermore, in low-shearing regions, such as post-stenotic recirculation regions, the hinge region of a mechanical heart valve or near the struts of a stent, RBCs tend to aggregate and form rouleaux microstructure that has significant viscoelastic properties~\cite{Fukada1980,Liu2006}. Unfortunately, few studies have examined arterial and cardiovascular device flows, including blood viscoelastic constitutive models (two of them, namely, Oldroyd-B and sPTT, are presented in section~\ref{section3}), compared with those available that consider blood's shear-thinning and viscoplastic behaviours. This is because generalised Newtonian models modify the momentum equation through a local, instantaneous apparent viscosity, whereas viscoelastic constitutive models introduce additional tensorial state variables whose evolution is governed by highly nonlinear constitutive equations. Consequently, viscoelastic simulations require additional degrees of freedom, robust treatment of stress advection, and often specialised numerical formulations such as log-conformation methods~\cite{Afonso2009}. These requirements become particularly demanding at high Deborah numbers, where elastic stresses can become large, and the coupled momentum-constitutive system may exhibit severe convergence difficulties~\cite{Alves2021}. Although some studies have shown significant differences in the flow field and in surface-averaged clinically important parameters, such as wall shear stress, between the Newtonian and viscoelastic models, those studies also accounted for shear-thinning behaviour alongside blood viscoelasticity~\cite{Chauhan2026}. Therefore, it is difficult to predict whether those differences arise solely from blood viscoelasticity, given that shear-thinning was also present. This necessitates further systematic investigation to determine the explicit effect of blood viscoelasticity on hemodynamics, even for simpler arterial flows, let alone the complex cardiovascular device flows.
    
    \item Likewise, the blood elasticity, the influence of blood plasticity or yield stress has not been investigated explicitly on the hemodynamics. However, blood indeed possesses a finite yield stress range between 0.0018-004~\cite{beris2021recent}. The viscoplastic models used to simulate blood flows, such as the Casson model, inherently include shear-thinning behaviour alongside blood yield stress. Therefore, once again, it is difficult to isolate the effect of each of these rheological behaviours on the hemodynamics, either in arterial flows or in cardiovascular device flows. The yield-stress behaviour of blood is expected to become important primarily in regions of the cardiovascular system where the local shear stress is sufficiently small to become comparable to the effective yield stress associated with red blood cell aggregation and rouleaux formation. Under sufficiently high shear, RBC aggregates are progressively disrupted, and blood behaves predominantly as a shear-thinning fluid with little sensitivity to yield-stress effects; consequently, the yield stress is unlikely to significantly influence bulk flow in large arteries under normal physiological conditions. Its importance can, however, increase substantially in low-shear, slowly recirculating, or nearly stagnant regions, where the ratio of yield stress to the characteristic viscous stress becomes appreciable. Such conditions can occur downstream of arterial stenoses, within aneurysmal sacs, near arterial bifurcations, and in separated-flow regions behind geometric obstructions, where local velocity gradients can become very small. The effect may also become transiently important during the low-flow portion of the cardiac cycle, particularly during late diastole, when regions that are fully yielded during systole may approach a weakly yielded or effectively unyielded state. Similar conditions are encountered in cardiovascular devices such as mechanical heart valves, ventricular-assist devices, blood pumps, and stents, where high-shear regions coexist with poorly washed cavities, hinge regions, wakes, and recirculation zones. In these low-shear regions, incorporating an effective yield stress can modify the size and structure of recirculation zones, reduce local velocities, increase residence times, and consequently influence hemodynamic quantities associated with thrombosis and blood stasis. A useful criterion for assessing its relevance is the ratio of the local yield stress to the characteristic local shear stress, $\tau_y/\tau_{\mathrm{local}}$: when $\tau_y/\tau_{\mathrm{local}}\ll1$, yield-stress effects are expected to be negligible, whereas when $\tau_y/\tau_{\mathrm{local}}\sim O(1)$, they can substantially influence the local flow. It should nevertheless be noted that the existence of a true, rate-independent yield stress in blood remains subject to debate because the apparent low-shear yield behaviour can also arise from RBC aggregation and the associated very large, time-dependent viscosity; therefore, the term effective or apparent yield stress is often more appropriate when interpreting Casson or Herschel-Bulkley descriptions of blood. Overall, yield-stress effects should be viewed as a local and flow-dependent phenomenon rather than a property that necessarily controls haemodynamics throughout an entire artery or cardiovascular device. 

    \item Another rheological behaviour of blood that has not been investigated so far, even for arterial flows, is the thixotropic behaviour. As discussed earlier in section~\ref{section2}, blood exhibits thixotropy, which arises mainly from the reversible aggregation and disaggregation of red blood cells. At low shear rates, RBCs form rouleaux and aggregates, increasing viscosity, whereas higher shear rates break these structures and reduce viscosity. When shear is reduced, the aggregates gradually reform, giving rise to time-dependent viscosity and hysteresis, a signature of blood thixotropy~\cite{thurston1979rheological,stoltz1981viscoelasticity,huang1995viscoelastic}. Blood thixotropy can be important in arterial and cardiovascular device flows because the shear rate varies continuously during the cardiac cycle, causing repeated aggregation and disaggregation of red blood cells. During low-shear phases, particularly diastole or in recirculation and stagnant regions of cardiovascular devices, RBC aggregation can develop and increase the apparent viscosity, whereas the high shear rates during systole progressively disrupt these aggregates. Because this structural evolution occurs over a finite timescale, the instantaneous blood viscosity depends not only on the current shear rate but also on its flow history. Consequently, thixotropy can influence the temporal evolution of wall shear stress, pressure drop, flow resistance, and regions of flow separation or stagnation, making it potentially important for accurate prediction of hemodynamics in arteries and blood-contacting cardiovascular devices. Despite this, no study has examined how the thixotropic behaviour of blood would influence hemodynamics in arterial and cardiovascular device flows compared with simple Newtonian or even generalised Newtonian fluid models. For the latter models, the blood apparent viscosity would be a function of local deformation rate, i.e., $\eta = \eta \left(\dot{\gamma}, \right)$, whereas in the case of the thixotropic model, it would be a function of shear rate and a structural parameter $\lambda$ that would consider the aggregation and disaggregation of RBCs, i.e., $\eta = \eta (\dot{\gamma}, \lambda)$. There is a greater scope for future research in this direction. 
\end{itemize}

Although the individual rheological characteristics of blood have been discussed separately here, it is important to note that these behaviours do not generally occur in isolation during physiological flow. Blood is a complex suspension of deformable erythrocytes, leukocytes, platelets, and plasma proteins, and its macroscopic rheological response emerges from the coupled dynamics of these constituents over a broad range of deformation rates and timescales. Consequently, when blood is subjected to a time-dependent deformation, several rheological mechanisms can act simultaneously. In particular, as mentioned earlier, shear-dependent viscosity may arise from the formation and disruption of red-blood-cell aggregates, yielding-like behaviour may become important when the applied stress is insufficient to overcome the strength of the microstructural network, while viscoelasticity reflects the finite relaxation of the cellular and plasma-protein microstructure. In addition, the reversible aggregation and disaggregation of erythrocytes introduces a time-dependent structural response, giving rise to thixotropy. Thus, viscoelasticity, viscoplasticity, and thixotropy should not necessarily be regarded as mutually exclusive mechanisms; rather, they can coexist and interact, particularly under pulsatile and spatially heterogeneous hemodynamic conditions. From this perspective, the thixo-elasto-viscoplastic (TEVP) model proposed by Spyridakis et al.~\cite{Spyridakis2023} provides a more comprehensive framework for representing the coupled rheological response of blood than models that account for only one or two of these mechanisms. The principal advantage of a TEVP description is that it allows the evolution of the blood microstructure and its mechanical response to be represented within a unified constitutive framework. This is particularly relevant to cardiovascular flows, where the local deformation history is continuously changing because of pulsatility, vessel curvature, branching, flow separation, and recirculation. The instantaneous rheological state of blood may therefore depend not only on the local shear rate but also on its previous deformation history. Such history dependence cannot be adequately captured by a purely viscous generalised Newtonian model, for which the stress is determined solely by the instantaneous rate of deformation. A TEVP model can, in principle, account for the coupled effects of structural evolution, elastic stress storage and relaxation, yielding, and shear-dependent viscous dissipation, thereby providing a more physically complete representation of blood rheology. Nevertheless, this additional physical fidelity comes at a substantial computational cost. Unlike Newtonian and generalised Newtonian formulations, TEVP models require solving additional transport and constitutive equations governing the evolution of structural and stress variables. The numerical treatment becomes particularly demanding in three-dimensional, pulsatile, patient-specific, or cardiovascular-device geometries, where strong spatial gradients, moving or complex boundaries, and multiple characteristic timescales can lead to additional challenges in convergence and stability.

If computational cost and numerical implementation were not limiting factors, a fully coupled TEVP formulation would arguably be the preferred modelling framework whenever the objective is to investigate the complete rheological influence of blood. However, despite its potential advantages, the TEVP framework has been employed in only a limited number of hemodynamic investigations to date~\cite{Giannokostas2023}. More importantly, its quantitative benefits relative to the much simpler Newtonian and generalised Newtonian descriptions have not yet been systematically established. A rigorous comparative assessment in which identical arterial or cardiovascular-device geometries, flow conditions, and boundary conditions are simulated using Newtonian, shear-thinning, viscoplastic, viscoelastic, and TEVP models would therefore be highly valuable. Such a study could determine under which physiological or pathological conditions the additional complexity of the TEVP model produces a meaningful change in predicted hemodynamic quantities. This would also help establish whether the inclusion of elasticity, yielding, and structural memory is essential for particular flow regimes or whether a simpler constitutive description is sufficient for engineering-level predictions. Experimental validation represents another important step toward establishing the relevance of the TEVP model to cardiovascular flows. In vitro experiments using anatomically realistic vascular geometries or cardiovascular-device replicas, as well as in vivo measurements where feasible, could be used to assess whether TEVP predictions provide a closer representation of physiological flow than conventional Newtonian or generalised Newtonian models. Similar validation strategies have previously been employed for generalised Newtonian descriptions of blood~\cite{Abbasian2020}. Importantly, such validation should not be restricted to global or surface-averaged quantities such as volumetric flow rate, pressure drop, or mean wall shear stress. While these quantities are useful, they can conceal substantial differences in the underlying local flow structure. A more stringent validation should therefore include spatially and temporally resolved quantities, particularly wall shear stress, velocity, and vorticity fields.

Wall shear stress is of particular interest because it represents the mechanical loading experienced by the vascular endothelium and is widely considered an important hemodynamic marker associated with cardiovascular disease~\cite{Cecchi2011,Malek1999}. Local regions of low, high, or oscillatory WSS can arise from flow separation, recirculation, stagnation, and complex vortical motion, and these features may influence endothelial function and the initiation or progression of vascular disease. Consequently, comparison of predicted WSS distributions could provide a substantially more sensitive assessment of different blood-rheology models than comparison of spatially averaged values alone. Furthermore, experimental measurements of local velocity and vorticity fields using four-dimensional flow MRI can provide a direct means of assessing the predicted three-dimensional and time-dependent flow structures~\cite{Soulat2020,Markl2012}. Such measurements can resolve flow acceleration, separation, recirculation, and coherent vortical structures, which in turn determine local velocity gradients and, hence, the distribution of WSS and other hemodynamic stresses. Similar spatially resolved measurements are also highly relevant to cardiovascular-device flows, where local flow disturbances, high-shear regions, and prolonged residence times can contribute to platelet activation, thrombosis, and blood damage. A comprehensive comparison between TEVP predictions and experimentally measured local flow fields would therefore provide a much stronger test of the physical relevance of blood-specific rheology. Such an approach could determine whether the inclusion of viscoelasticity and thixotropy yields measurable changes in the flow structures ultimately relevant to disease development and device performance. This is particularly important because blood rheology has already been implicated as an important factor in several cardiovascular and hematological diseases~\cite{Lowe1986,nader2019blood}. Nevertheless, the specific contributions of viscoelasticity, viscoplasticity, and thixotropic structural memory to the onset and progression of cardiovascular disease remain considerably less well established than the effects of shear-dependent viscosity. Establishing these contributions will require a coordinated combination of constitutive modelling, high-fidelity numerical simulations, controlled in vitro experiments, and, where possible, in vivo measurements. Such investigations could ultimately clarify when the additional complexity of a TEVP description is physically warranted and whether incorporating the complete rheological response of blood can yield clinically or technologically relevant insights into disease progression, cardiovascular device performance, and therapeutic design. 

\section*{Data availability statement}
No new data were created or analysed in this study.

\section*{Conflict of interest}
The authors declare that there are no competing interests.

\section*{Funding}
The author would like to thank the Science and Engineering Research Board (SERB), Government of India (Project no. ECR/2018/000202) and the Indian Council of Medical Research (ICMR), Government of India (Project no. 17X(3)/Adhoc/17/2022-ITR) for providing the necessary funding to carry out the research work.   

\printnomenclature
\bibliography{mybibfile}

\end{document}